\documentclass[sigconf]{acmart}

\usepackage{booktabs} 
\usepackage{natbib}

\setcopyright{none}

\usepackage{microtype}

\usepackage{multirow}
\usepackage{booktabs} 
\usepackage{amsfonts}
\usepackage{amssymb}
\usepackage{graphicx}
\usepackage{float}
\usepackage{siunitx}
\usepackage{tikz} 
\usepackage{pgfplots}
\usepackage{mathtools}
\usepackage{amsmath}
\usepackage{amssymb}
\usepackage{subcaption}
\usepackage{makecell}
\usepackage{float}
\usepackage{pifont}
\newcommand{\cmark}{\ding{51}}%
\newcommand{\xmark}{\ding{55}}%
\pgfplotsset{compat=newest} 
\usepgfplotslibrary{units} 
\usepackage{hyperref}
\usepackage{booktabs} 
\usepackage{colortbl} 
\usepackage{xcolor} 
\usepackage{xfrac}
\usepackage{enumitem} 
\usepackage{booktabs} 
\usepackage{algorithm} 
\usepackage{algpseudocode}
\usepackage{listings}
\usepackage{color}
\usepackage[normalem]{ulem}
\newcommand{\pevbis}[1]{{#1}}

\newcommand{\maria}[1]{{#1}}
\usepackage{pgfplots}
\definecolor{mygreen}{rgb}{0,0.6,0}
\definecolor{mygray}{rgb}{0.5,0.5,0.5}
\definecolor{mymauve}{rgb}{0.58,0,0.82}

\lstdefinestyle{customc}{
	belowcaptionskip=1\baselineskip,
	breaklines=true,
	xleftmargin=\parindent,
	language=C,
	showstringspaces=false,
	basicstyle=\scriptsize\ttfamily,
	keywordstyle=\bfseries\color{green!40!black},
	commentstyle=\itshape\color{purple!40!black},
	identifierstyle=\color{blue},
	stringstyle=\color{orange},
}
\lstdefinestyle{customasm}{
	belowcaptionskip=1\baselineskip,
	xleftmargin=\parindent,
	language=[x86masm]Assembler,
	basicstyle=\scriptsize\ttfamily,
	commentstyle=\itshape\color{purple!40!black},
}
\usepackage{amssymb}
\usepackage{bm}

\usepackage{enumitem}

\setcopyright{none}
\renewcommand\footnotetextcopyrightpermission[1]{}
\begin{document}
	\fancyhead{}
	\title{Defeating Opaque Predicates Statically through Machine Learning and Binary Analysis}
	
	\author{Ramtine Tofighi-Shirazi}
	\affiliation{
		\institution{Univ. Grenoble Alpes, CNRS, Institut Fourier}
		\institution{Trusted Labs, Thales Group}
		\streetaddress{6 rue de la Verrerire}
		\city{Meudon}
		\state{France}
	}
	\email{ramtine.tofighishirazi@thalesgroup.com}

		\author{Irina M\u{a}riuca As\u{a}voae}
	\affiliation{
		\institution{Trusted Labs, Thales Group}
		\streetaddress{6 rue de la Verrerire}
		\city{Meudon}
		\state{France}
	}
	\email{irina-mariuca.asavoae@thalesgroup.com}
	
		\author{Philippe Elbaz-Vincent}
	\affiliation{
		\institution{Univ. Grenoble Alpes, CNRS, Institut Fourier}
		\streetaddress{100 Rue des Mathematiques}
		\city{F-38000 Grenoble}
		\state{France}
	}
	\email{philippe.elbaz-vincent@univ-grenoble-alpes.fr}

	\author{Thanh-Ha Le}
	\affiliation{
		\institution{Work done when author was at Trusted Labs}
		\streetaddress{6 rue de la Verrerire}
		\city{Meudon}
		\state{France}
	}
	\email{lethanhha.work@gmail.com}

	\renewcommand{\shortauthors}{Ramtine Tofighi-Shirazi \textit{et al.}}

	\begin{abstract}
		We present a new approach that bridges binary analysis techniques with machine learning classification for the purpose of providing a static and generic evaluation technique for opaque predicates, regardless of their constructions.
		We use this technique as a \textit{static automated deobfuscation tool} to remove the opaque predicates introduced by obfuscation mechanisms.
		According to our experimental results, our models have up to 98\% accuracy at detecting and deobfuscating state-of-the-art opaque predicates patterns. 
		By contrast, the leading edge deobfuscation methods based on symbolic execution show less accuracy mostly due to the SMT solvers constraints and the lack of scalability of dynamic symbolic analyses.
		Our approach underlines the efficiency of hybrid symbolic analysis and machine learning techniques for a static and generic deobfuscation methodology.
	\end{abstract}

	%
	%

	\keywords{Obfuscation, deobfuscation, machine learning, symbolic execution, opaque predicate}

	\maketitle

\section{Introduction}
	Automatic program analysis is widely used in research and industries for various software evaluation purposes. 
	In this context, software reverse engineering, which consists in the understanding of the internal behavior of a program, relies on various program analyses techniques such as static or dynamic symbolic execution. 
	To prevent the application of software reverse engineering techniques, code obfuscation~\cite{Ctaxonomy, DBLP:conf/ccs/SutterBCFZWD16} is a broadly employed protection methodology which transforms a program into an equivalent one that is more difficult to understand and analyze. 
	Among these obfuscation mechanisms, opaque predicates represent one of the major and fundamental obfuscation transformations used by obfuscators to mitigate the risk of reverse engineering. 
	Opaque predicates represent constant branching conditions that are obfuscated for the purpose of hiding the fact that they are constant. 
	Thus, an opaque predicate value is known a-priori by the defender, but shall be hard to deduce for an attacker. 
	We choose this obfuscation transformation for its variety of types and constructions and their common use as foundation for other obfuscation transformations as means of improving the transformations robustness and resiliency. 
	Opaque predicates~\cite{CollbergTL98} are widely used as technique for various security applications, \textit{e.g.} metamorphic malware mutation~\cite{DBLP:conf/dimva/BruschiMM06}, Android applications~\cite{DBLP:conf/iait/Kovacheva13} or white-box cryptographic implementations. 
	As a consequence, several works focus on the deobfuscation of opaque predicates (\textit{e.g.}~\cite{DBLP:conf/ccs/EyrollesGV16, DBLP:conf/ccs/XWW15, DBLP:conf/sp/BardinDM17, DBLP:conf/amast/PredaMBG06, DBLP:conf/uss/BlazytkoCAH17, DBLP:conf/acsac/Tofighi-Shirazi18, DBLP:journals/compsec/BiondiJLS17}) in order to evaluate the quality of the obfuscated code rendered by this transformation. 
	However, these techniques are often based on dynamic analysis and are therefore limited or not scalable.
	
	\paragraph{\textbf{Problem setting:}}
	Existing state-of-the-art opaque predicates deobfuscation techniques and tools suffer from the following limitations:
	~\\
	\begin{enumerate}
		\item \textit{Specificity:} Techniques that evaluate opaque predicates are often focused on specific constructions, hence lacking of generality towards all existing patterns of such obfuscation transformation.
		\item \textit{Code coverage:} Most recent deobfuscation techniques are based on dynamic symbolic execution which require the generation of instruction traces. The ability to cover all paths of the program is an issue that prevents, in some cases, the complete code deobfuscation.
		\item \textit{Scalability:} Dynamic symbolic execution techniques may also lack of scalability on some types of code 
		such as malwares that use specific triggers (\textit{e.g.} an input from a Command and Control server) to execute 
		the non-benign part of the code. Thus, dynamic analysis may not scale and cover the non-triggered part of the code.
		\item \textit{Satisfiability modulo theories solvers:} SMT solvers used in path-reachability analyses suffer from several limitations depending on the constructions of the opaques predicates. Some constructions that are based on aliases or mixed boolean and arithmetic expressions usually cause SMT solvers to fail at predicting the feasibility of a path.
	\end{enumerate}
	~\\
	 Our work has the goal to re-introduce static analysis for obfuscated software evaluation and deobfuscation. To this end, we propose a new approach that bridges static symbolic execution and machine learning models to provide a generic evaluation of opaque predicates.

	We present several studies and experiments towards the construction of machine learning models that can either detect an opaque predicate or predict its invariant value without executing the code.
	We also extend our design to the deobfuscation of such obfuscation transforms, regardless of their constructions, by creating a static analysis plug-in within a widely used reverse engineering tool called IDA~\cite{IDAPro}.
	To further evaluate our methodology, we compare it against available static and dynamic symbolic-based tools for the deobfuscation of opaque predicates. 
	We conduct further evaluations against obfuscators such as Tigress~\cite{Tigress} and OLLVM~\cite{ieeespro2015-JunodRWM}.

	The aftermath of our work shows that combining machine learning techniques with static symbolic analysis provides a generic, automatic, and accurate methodology towards the evaluation of opaque predicates. Our work shows that machine learning enables a better efficiency and genericity for this application, while allowing us to work without SMT solvers to predict reachable paths.
~\\

\paragraph{\textbf{Contributions:}}
\begin{enumerate}[leftmargin=*]
	\item We present our novel methodology that binds binary analysis and machine learning techniques to evaluate and deobfuscate opaque predicates statically.
	A presentation of several studies towards an efficient and accurate creation of machine learning models is also given.
	\item The evaluation of our methodology against state-of-the-art obfuscators such as Tigress and OLLVM, as well as novel opaque predicate constructions such as \textit{bi-opaque} predicates.
	\item The illustration of the efficiency of our methodology, used as a static analysis deobfuscation tool, on several datasets by comparing it to existing state-of-the-art deobfuscation tools based on symbolic execution and SMT solvers.
\end{enumerate}
~\\
Our paper is organized as follows: in Section \ref{sec:background} we recall background information on opaque predicates types, constructions,  and deobfuscation methods. Then we introduce some notions of supervised machine learning. 
In Section \ref{sec:design}, we present our methodology which combines binary analysis and machine learning to achieve an efficient evaluation and deobfuscation of opaque predicates. Section \ref{sec:experiments} presents our experiments towards an accurate model construction, whereas Section \ref{sec:evals} illustrates our evaluations on state-of-the-art publicly available obfuscators.
A comparison to existing symbolic-based deobfuscation techniques against our methodology is also provided in Section \ref{subsec:deobf}.
We then describe our design limitations and perspectives in Section \ref{sec:limitations}, along with related work in Section \ref{sec:relatedwork}, before concluding in Section \ref{sec:conclusion}.

\section{Technical Background}
\label{sec:background}
\pevbis{In this section we briefly present opaque predicates, obfuscation and deobfuscation techniques, and introduce several notions related to supervised machine learning.}

\subsection{Code obfuscation}
	Collberg \textit{et al.}~\cite{Ctaxonomy} define code obfuscation as follows: 
	\\Let $P \xrightarrow{T} P'$ be a transformation $T$ of a source program $P$ into a target program $P'$. We call $P \xrightarrow{T} P'$ an
	\textit{obfuscating transformation} if $P$ and $P'$ have the same observable behavior. 	
	Consequently, the following conditions must be fulfilled for an obfuscating transformation : if $P$ fails to terminate, or terminates 
	with an error condition, then $P'$ may or may not terminate; otherwise, $P'$ must terminate and produce the same output as $P$.
	Several obfuscation transformations exist, each of them having their own purpose: obfuscate the layout, the data, or the control-flow of a program. 
	A classification of all these obfuscations, as well as known deobfuscation methods with their different purposes, has been provided by S. Schrittwieser \textit{et al.}~\cite{DBLP:journals/csur/SchrittwieserKK16}.

\subsection{Opaque predicates}
Often combined with bogus code, opaque predicates~\cite{CollbergTL98} aim at encumbering control-flow graphs with redundant infeasible paths. 
Compared to other control-flow obfuscation transformations such as control-flow flattening or call/stack tampering~\cite{DBLP:journals/tse/LakhotiaKV05}, opaque predicates  are supposedly more stealthy (\textit{i.e.} hard for an attacker to detect) because of the difficulty to differentiate an opaque predicate from original path conditions in a binary code. 
In the followings, we give an overview of some existing types and constructions of opaque predicates.

\subsubsection{Opaque predicate types.}
We denote by $\phi$ a predicate, \textit{i.e.} a conditional jump, within a binary code. 
Such predicate can be evaluated to both \textit{true} or \textit{false} (\textit{i.e.} $0$ or $1$). 
We denote by $\mathcal{O}$ the obfuscation function that generates opaque predicates.
$\mathcal{O}$ takes as input a predicate $\phi$ and produces the opaque predicate $\mathcal{O}(\phi)$.
For security purposes, $\mathcal{O}(\phi)$ should be stealthy (indistinguishable from any other $\phi$) and resilient (its value should not be easily known by an attacker).
There are two types of opaque predicates, namely the \textit{invariant} and the \textit{two-ways}.
C. Collberg \textit{et al.} define these predicates by, respectively, $P^T$, $P^F$, and $P^?$.
\maria{Our methodology aims at detecting and deobfuscating opaque predicates of types $P^T$ and $P^F$.} 
Next we explain the introduced notations $P^T$, $P^F$ and $P^?$.

\paragraph{\textbf{Invariant opaque predicates:}}
Let $\mathcal{O}(\phi) : X \rightarrow \{0, 1\}$ be an obfuscated predicate that evaluates to either 0 or 1 and $\mathcal{O}$ be the function that obfuscates the predicate. 
We denote by $X$ the set of all possible inputs. If for all $x \in X$, $\mathcal{O}(\phi)(x) = 1$ (resp. $0$), then we say that the 
predicate is \textbf{always true} (resp. \textbf{always false}), denoted $P^T$ (resp. $P^F$).
These opaque predicates are said \textit{invariant}, as they always evaluate to the same value for all possible inputs.

\paragraph{\textbf{Two-ways opaque predicates:}}
	Another type of opaque predicates is referred to as \textit{two-way}, which can evaluate to both true or false for all possible inputs. 
	Such a construction requires both branches to be semantically equivalent in order to preserve the functionality of the code that 
	will be executed.
	In other words we have, if for all input $x \in X$, $Pr_{x\leftarrow X}[\mathcal{O}(\phi)(x) = 1] = \frac{1}{n}$ with $n \in \mathbb{N}^{+}$, then the predicate is \textbf{either true or false}, regardless of $x$.

\subsubsection{Opaque predicate constructions.}
\label{paragraph:opaqueconstruct}
Several proposals exist in the literature about how to construct a resilient and stealthy opaque predicate, \textit{e.g.} \cite{DBLP:journals/ecr/MylesC06, DBLP:conf/dsn/XuZKTL18}.
Each of these constructions has the purpose to thwart specific deobfuscation analyses and will be summarized
in Section \ref{subsec:deobfback}.

	\paragraph{\textbf{Arithmetic-based opaque predicates}} Constructed using mathematical formulas which are hard to solve, they aim at encoding invariants into arithmetic properties on numbers. 
	
	\paragraph{\textbf{Mixed-boolean and arithmetic based opaque predicates}} Otherwise known as MBA~\cite{DBLP:conf/wisa/ZhouMGJ07}, they are based on a data encoding technique using linear identities involving boolean and arithmetic operations.
	The resulting encoding is made dependent on external inputs such that it cannot be deobfuscated by compiler optimizations.
	
	\paragraph{\textbf{Alias-based opaque predicates}} They are one of the first choices of Collberg \textit{et al.}~\cite{CollbergTL98} for their construction.
	Aliasing is represented by a state of a program where certain memory location is referenced to by multiple pointers, and pointer alias analysis is undecidable.
	
	\paragraph{\textbf{Environment-based opaque predicates}} These constructions use constant elements lifted from the system, or libraries. 
	
	\paragraph{\textbf{Bi-opaque opaque predicates}} 
	Bi-opaque constructs aim at employing the weaknesses of symbolic execution, and are introduced in recent work~\cite{DBLP:conf/dsn/XuZKTL18}. 
	Based on the observation that deobfuscation techniques using symbolic execution are prone to some challenges and limitations, bi-opaque predicates intend to either introduce false negatives or false positives results. Such construction has been shown effective against state-of-the-art deobfuscation 
	tools based on dynamic symbolic execution, such as Triton~\cite{SSTIC2015-Saudel-Salwan} or Angr~\cite{shoshitaishvili2016state}.
\subsection{Deobfuscation}
\label{subsec:deobfback}
Due to their wide utilization and popularity, opaque predicates are target of several published attacks. 
Each of these deobfuscation methodologies has strengths and limitations as summarized in Table \ref{fig:constructvsdeobf}.

A first deobfuscation technique called \textit{probabilistic check} consists in executing $n$ times a program segment to see if a 
predicate is invariant.
Such technique can be combined with fuzzing on the inputs. However, it is prone to high false positives and negatives results while depending on the possibility to execute $n$ times the code.

Also, due to the overhead introduced by most complex opaque predicates constructs, it has been showed in the literature that 
there are surprisingly relatively 
few predicates that are used over and over again.
This leads to a possible \textit{pattern matching attack} (\textit{i.e.} dictionary attack)~\cite{DBLP:conf/ccs/EyrollesGV16}, where one takes obfuscated predicates from a program being attacked and pattern-matches the source code against known examples. 
Nevertheless, it is possible to build variants of opaque predicates that cannot be matched using dictionary attacks, which implies a high false negative rate.

Another technique that uses \textit{abstract interpretation}~\cite{DBLP:conf/amast/PredaMBG06} provides correctness and
efficiency in the deobfuscation of some specific constructions of opaques predicates.
It consists in a static symbolic attack that can be only efficient against some classes of invariant arithmetic-based 
opaque predicates, but does not focus on other types and structures.

Another recently introduced technique~\cite{DBLP:conf/uss/BlazytkoCAH17} uses \textit{program synthesis}.
Originally designed for the deobfuscation of virtualized code, this approach has been successful for the simplification of MBA expressions. 
\begin{table*}[h!]
	\centering
	\resizebox{\textwidth}{!}{  
		\begin{tabular}{|c|c|c|c|c|c|}
			\hline 
			\rowcolor{cyan!10}\textbf{Constructions} & \textbf{Probabilistic check} & \textbf{Pattern matching} & \textbf{Abstract interpretation} & \textbf{Automated proving} & \textbf{Program synthesis} \\ 
			
			\hline 
			\rowcolor{black!10}\textbf{Arithmetic-based} & {\colorbox{black!10}{\makecell{\cmark\cite{DBLP:conf/wcre/UdupaDM05}\\High FN/FP}}} & {\colorbox{black!10}{\makecell{\xmark\\High FN}}} & \cmark\cite{DBLP:conf/amast/PredaMBG06} & \cmark\cite{DBLP:conf/ccs/XWW15, DBLP:conf/sp/BardinDM17} & \xmark \\

			\textbf{MBA-based} & \makecell{\xmark\\High FN/FP} & \cmark\cite{DBLP:conf/ccs/EyrollesGV16} & \xmark & \makecell{\xmark\\(limitations of SMT solver)} & \cmark\cite{DBLP:journals/compsec/BiondiJLS17, DBLP:conf/uss/BlazytkoCAH17} \\

			\rowcolor{black!10}\textbf{Alias-based} & {\colorbox{black!10}{\makecell{\xmark\\High FN/FP}}} & \xmark & \xmark & {\colorbox{black!10}{\makecell{\xmark\\(limitations of symbolic execution)}}} & \xmark \\

			\rowcolor{black!10}\textbf{Environment-based} & {\colorbox{black!10}{\makecell{\xmark\\High FN/FP}}} & \xmark & \xmark & \cmark\cite{DBLP:conf/ccs/XWW15, DBLP:conf/sp/BardinDM17} & \xmark \\

			\textbf{Bi-opaque} & \makecell{\xmark\\High FN/FP} & \xmark & \xmark & \makecell{\xmark\\High FN/FP} & \xmark \\ 
			
			\hline 
		\end{tabular}
	}
	
	\caption{Illustrations of opaque predicates deobfuscation strengths and targets.} 
	\label{fig:constructvsdeobf}
\end{table*}

Moreover, current state-of-the-art deobfuscation approaches use \textit{automated proving} to compute if a predicate is opaque~\cite{DBLP:conf/ccs/XWW15, DBLP:conf/sp/BardinDM17}.

	Udupa \textit{et al.}~\cite{DBLP:conf/wcre/UdupaDM05} use static path feasibility analysis based on SMT solvers to determine the reachability of execution paths. 
	Their methodology inherit the limitations of static analysis, namely path explosion.
	This is the reason why recent automated proving techniques are based on dynamic symbolic execution (\textit{i.e.} DSE) to check path feasibility or infeasibility~\cite{DBLP:conf/sp/BardinDM17} and remove opaque predicates.  
	Yet, automated proving based analysis, either static or dynamic, suffers from the SMT solvers restrictions. 
	It has been showed that SMT solvers fail against MBA opaque predicates, alias-based constructions, and can even be misguided by more recent constructions such as bi-opaque predicates.

	Overall, DSE is currently considered the most effective methodology against opaque predicates, but the evaluation of such technique 
	has been shown effective mainly against arithmetic or environment based opaque predicates.
	This demonstrates the importance of a generic and scalable methodology that can evaluate both stealth and resilience of opaque predicates
	for all existing constructions.

\subsection{Supervised Machine Learning}
The decision of labeling a predicate as opaque, and even more as invariant $P^T$ or $P^F$ opaque predicate, can be considered as classifications problems.
Our target is to find algorithms that work from external supplied instances (\textit{e.g.} binaries, instructions traces, etc.) in order to produce general hypotheses. 
From these hypotheses, we want to make predictions about future instances. 
\textit{Supervised machine learning} provides a dedicated methodology that achieves this goal.
The aim of a supervised machine learning is to build a \textit{classification model} which will be used to assign \textit{labels} to testing or unknown instances.  
In other words, let $X$ be our inputs (\textit{i.e.} instances) and $Y$ the outputs (\textit{i.e.} predicted labels). 
A supervised machine learning algorithm will be used to learn the mapping function $f$ such that $Y = f(X)$.
The goal is to approximate $f$ such that for any new instance $X$ we can predict its label $Y$. 
In our case the inputs are represented by $n$-dimensional vectors of numerical features that represent these features, \textit{i.e. features vectors}, for which the extraction is described in the following paragraph.

\subsubsection{Feature extraction.}
In the machine learning terminology, the inputs of a model are usually derived from what is called \textit{raw data}, \textit{i.e.} the data samples we want to classify or predict. 
These data samples cannot be directly given to a classification model and need to be processed beforehand. 
This processing step is called \textit{feature extraction} and consists in combining the raw data variables into numerical features.
It allows to effectively reduce the amount of data that must be processed, while accurately describing the original dataset of raw data.
In our case, since raw data are text documents (\textit{e.g.} disassembly code, symbolic execution state, etc.), one practical use of feature extraction consists in extracting the \textit{words} (\textit{i.e.} the features) from the raw data and classify them by frequency of use (\textit{i.e.} weights). 
Different approaches exist for understanding what a word is and to compute its weight.
In this paper we use the \textit{bag of words} approach which identifies terms with words. 
As for the weights, we studied \textit{term frequency} (\textit{i.e.} how frequently a word occurs in a document) with and without \textit{inverse document frequency}~\cite{DBLP:journals/jd/Jones04} used in Section \ref{sec:experiments} 
in order to select the best possible extraction technique. 

\subsubsection{Classification algorithms.}
The choice of which specific learning algorithm to use is a critical step.
Many classification algorithms exist~\cite{James:1985:CA:7557}, each of them having different mapping functions. 
Classification is a common application of machine learning. As such, there are many metrics that can be used to measure and evaluate 
the accuracy as well as the efficiency of our models.
In order to compute these metrics, \textit{$k$-Fold Cross-Validation} is a frequent technique~\cite{DBLP:conf/ijcai/Kohavi95}. 
It consists in reserving a particular set of samples on which the model does not train. 
This limited set of samples allows to estimate how the model is expected to perform on data not used during the training phase. 
The parameter $k$ refers to the number of groups that a given dataset of samples is split into, in order to calculate the mean of our models \textit{accuracy} as well 
as the \textit{F1 score} based on the value of $k$.
While the accuracy of the model represents the ratio of correctly predicted labels to the the total of labels, F1 score takes both false positives and negatives into account.
In our experimentations and evaluations, the accuracies and F1 scores are calculated using \textit{k-fold cross-validation}, with $k=20$ for a better generalization of our model to unknown instances.

\section{Our Methodology}
\label{sec:design}
\begin{figure*}[h!]
	\centering
	\makebox[\textwidth]{\includegraphics[width=\textwidth]{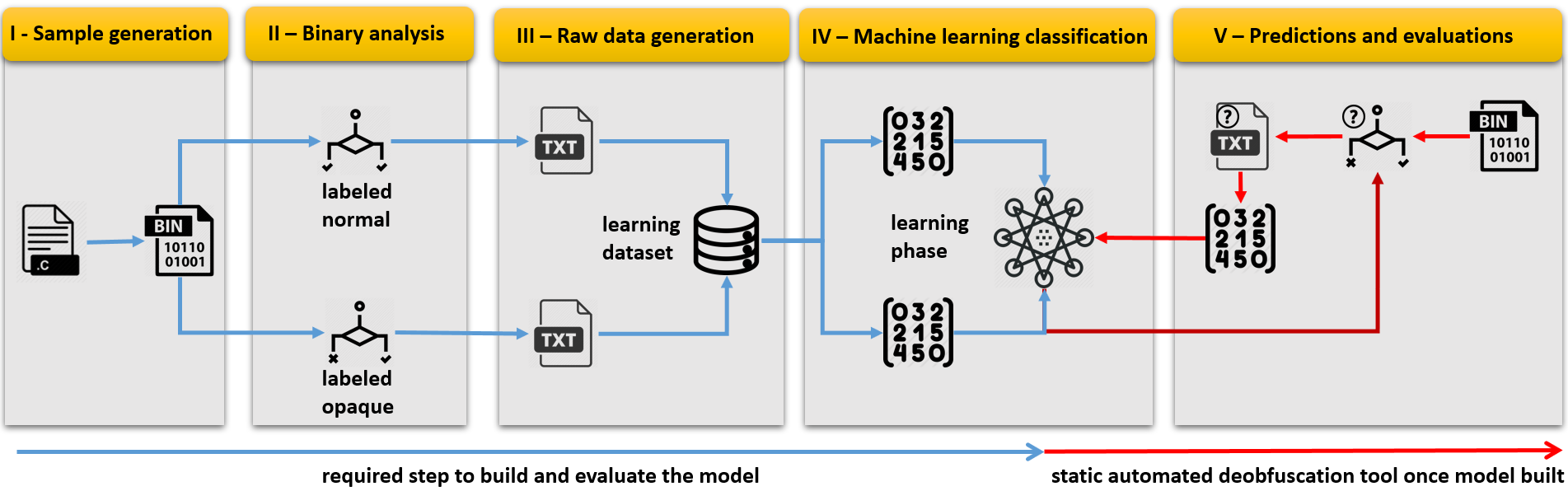}}
	\caption{Evaluation steps of opaque predicates.}
	\label{fig:design}	
\end{figure*}

Our methodology design is built in two parts.
	The first part consists in creating a machine learning model for the evaluation and deobfuscation of opaque predicates.
	The second part uses the validated model in order to remove such obfuscation transformation statically. 
	Figure \ref{fig:design} illustrates our methodology.
	The first step consists in generating a set of obfuscated binaries.
	Our datasets of C code samples are presented in Section \ref{subsec:datasets}.
	In the second step, the binary is disassembled and we collect and labelize each predicate, \textit{e.g.} defining if the predicate is opaque or normal, as described in Section \ref{subsec:binanalysis}. .
	The third step consists in a depth-first search algorithm to collect each path leading to a predicate. We use a thresholded static symbolic execution to collect our raw data for the machine learning model.
	These data are normalized, processed and used to train and validate our model in a fourth step, as presented in Section \ref{subsec:mlanalysis}. Finally, the fifth and final step shows that our model can be used and integrated in a static deobfuscation tool to predict and remove opaque predicates as presented in Section \ref{sec:evals}.

\subsection{Binary analysis}\label{subsec:binanalysis}
Our methodology relies on static symbolic execution to retrieve \textit{the semantics of the predicate constructions} before the machine learning classification models evaluates them.
Thus, a first step in our design is the generation of \textit{raw data}. 
This refers to a representation of data samples that contain noisy features and need to be processed in order to 
extract informative characteristics from the data samples, before training a model. 
Since our goal is to evaluate the opaque predicates, we choose to generate our raw data from the disassembled binary code control-flow graph.

Moreover, in order to have a \textit{scalable methodology}, we work statically in order to prevent the need of executing the code. 
	This approach also permits a \textit{better code coverage} compared to existing dynamic approaches.
	However, our approach can be extended to instruction traces in cases where the analyzed code is encrypted or packed.
The raw data used contains the symbolic expressions $\mathcal{S}$ of collected predicates $\phi$ denoted by $\mathcal{S}_{\phi}$.

We studied different formats and contents of such raw data as well as their impact on the efficiency of the trained model (see Section \ref{sec:experiments}).
In the following sections we present  the binary analysis part of our design, namely \emph{thresholded static symbolic execution}, which we employ to generate the raw data from predicates.

\subsubsection{Thresholded Static Symbolic Execution.}
Static symbolic execution is a binary analysis technique that captures the semantics (\textit{i.e.} logic) of a program. 
An interpreter is used to trace the program, while assuming symbolic values for inputs rather than obtaining concrete values as a normal execution would. 
A symbolic state $\mathbb{S}$ is built and consists in a set of symbolic expressions $S$ for each variables (\textit{i.e.} registers, memory, flags, etc.).
Several techniques exist for symbolic execution~\cite{DBLP:journals/csur/BaldoniCDDF18}.

In our work we use disassembled functions to collect the symbolic expressions of a predicate $\mathcal{S}_{\phi}$. 
We start by generating all possible paths from a function entry point to a predicate $\phi$ using a depth-first search algorithm. The latter prevents us from using SMT solvers to generate all feasible paths since they are prone to limitations and errors depending on the protections applied.
In order to avoid path explosion, we use a \textit{thresholded} static symbolic execution technique that bounds the number of paths generated for one predicate and the amount of time the analysis has to iterate on a loop.
Note that our methodology is intra-procedural since publicly available obfuscators, \textit{e.g.} Tigress, O-LLVM, generate intra-procedural opaque predicates.

\paragraph{\textbf{Path generation:}}
We denote by $\phi_{n}$ the $n$-th predicate within a disassembled function $F$ in a binary $B$. 
When a predicate is identified, we generate all paths from $F$ entry point to the collected $\phi_{n}$ using a depth-first search (\textit{i.e.} DFS) algorithm. 
DFS expands a path as much as possible before backtracking to the deepest unexplored branch. 
This algorithm is often used when memory usage is at a premium, however it remains hampered by paths containing loops.
Thus, we use two distinct thresholds, one for loop iterations denoted by $\alpha_{loop}$, and one for the number of paths to be generated denoted $\alpha_{paths}$. 

\paragraph{\textbf{Symbolic state generation:}}
In order to have a symbolic state, we use all collected paths of a predicate.
We denote by $\mathcal{P}$ the set of all collected paths $\sigma$ of a predicate $\phi$.
Let $\mathrm{S}$ be the symbolic execution interpreter function such that $\mathrm{S}(\sigma_{i}) = \mathbb{S}_{\phi^{\sigma_{i}}}$.
In other words, the symbolic execution interpreter $\mathrm{S}$ returns a symbolic state  $\mathbb{S}_{\phi^{\sigma_{i}}}$ for a path $\sigma_{i}$, $i \in [O, |\mathcal{P}|]$ 
of a predicate $\phi$.
The generated symbolic states for all predicates will be used as raw data and then be processed for the classification models.

\subsection{Machine learning}\label{subsec:mlanalysis}

We experiment different instances for our classification models to study the impact on their accuracy. 	
Since symbolic execution is often based on an intermediate representation that captures all the semantics as well as side effects of the assembly instructions, 
several intermediate representations exist and are widely used, \textit{e.g.} LLVM-IR or Miasm-IR~\cite{Miasm}.
We implemented our methodology using Miasm2 reverse engineering framework, which integrates translators from Miasm-IR to other languages (\textit{e.g.} SMT-LIBv2~\cite{BarFT-RR-17}, Python~\cite{Rossum:1995:PRM:869369}, or C~\cite{Kernighan:1988:CPL:576122}). 
This gives us the ability to study the impact of the language used to express the symbolic expressions, within our raw data, on our classification models.

\subsubsection{Raw data.}
\label{subsec:raw_data}
Intermediate representations use concrete values within their generated expressions. 
This causes raw data to depend on addresses that are specific to some binaries and prevents our models to scale on unknown data. 
Listing 1 illustrates this issue with one predicate symbolic expression in the Miasm-IR language.
Moreover, some intermediate representations, \textit{e.g.} Miasm-IR, use identifiers in order to express modified registers name or memory locations. This may further affect the scalability of our trained models.

For the purpose of having a model that can scale to unknown data, we use a normalization phase that replaces identifiers and concrete values by symbols, and non-alphanumerical characters by alphanumerical words. 
This is a necessary step for a complete features extraction phase that sometimes excludes non-alphanumerical characters when working on text-based raw data. In Listing 1 lines 10 and 13 we provide examples of the normalization step.

Since our methodology computes a full symbolic state from any function entry-point to a targeted predicate, there is a need to know if all information within the collected symbolic state is relevant for our models. 
The goal is to have many features for an accurate classification without adding too much noise.

\begin{lstlisting}[language=Python,caption={Miasm-IR predicates expressions before and after normalization}\label{lst:expr_addr}]
# Miasm-IR predicate expression before normalization
# Miasm-IR predicate expression of an P^T opaque predicate
ExprId('IRDst', size=64) = ExprInt(0x402b36, 64)

# Miasm-IR predicate expression of an P^F opaque predicate
ExprId('IRDst', size=64) = ExprInt(0x402209, 64)

# Miasm-IR predicate expression after normalization
# Normalized Miasm-IR predicate expression of an P^T opaque predicate
ExprId(id1, size=64) = ExprInt(v1, 64)

# Normalized Miasm-IR predicate expression of an P^F opaque predicate
ExprId(id1, size=64) = ExprInt(v1, 64)
\end{lstlisting}

Another issue to be avoided is having raw data samples that do not contain enough information to distinguish between samples that have different labels, as illustrated also in Listing 1.
In other words, we may have two expressions that are identical but have different labels, \textit{e.g.} the first being the expression of a $P^T$ and the second an expression of a $P^F$.
To avoid this matter we use the thresholded symbolic execution, which generates expressions for each path leading to a predicate.
Listing 2 illustrates the predicates expressions from Listing 1 along with others memory and registers expressions from their symbolic state. 
Namely, line 10 in Listing 1 corresponds to lines 2-5 in Listing 2, while line 13 in Listing 1 corresponds to lines 8-11 in Listing 2.
We can see that now we have more informations that allows us to distinguish between both predicates.

\begin{lstlisting}[language=Python, caption={Miasm-IR predicate expressions after our normalization phase}\label{lst:set3}]
# Normalized Miasm-IR predicate expression of an P^T opaque predicate
ExprId(id1, size=64) = ExprInt(v1, 64)
ExprId(id2, size=1) = ExprInt(v2, 1)
...
ExprId(id9, size=1) = ExprInt(v5, 1)

# Normalized Miasm-IR predicate expression of an P^F opaque predicate
ExprId(id1, size=64) = ExprInt(v1, 64)
ExprId(id2, size=1) = ExprCond(ExprMem(ExprId(v3, size=64), size=8), ExprInt(v4, 1), ExprInt(v3, 1))
...
ExprId(id9, size=1) = ExprInt(v3, 1)
\end{lstlisting}

We study the use of several expressions in our raw data to distinguish between sample that have different labels. 
To this end, we divide our instances into three sets:
~\\

\begin{itemize}
	\item \textbf{Set 1}: with samples containing only the expression of the predicate in a static single assignment form (\textit{i.e.} SSA) as illustrated in Listing 1.
	\item \textbf{Set 2}: with samples containing only the expressions of the predicate and its corresponding flags in a SSA form.
	\item \textbf{Set 3}: with samples containing the full symbolic state of a path, from an entry-point to a targeted predicate, \textit{i.e.} all memory, flags, and registers modified in a SSA form as illustrated in Listing 2.
\end{itemize}
~\\
In Section \ref{subsub:s2}, each set is studied in order to find the best possible raw data content.
We start by calculating for each set the similarity percentages based on 5000 samples of predicates, either \textit{normal} or \textit{opaque} predicates generated by the Tigress obfuscator on a dataset of C code samples (see Section \ref{subsec:datasets}). 
In other words, we search for raw data with different labels (\textit{e.g.} $P^F$ and $P^T$) but with the same content.
As we can see in Table \ref{fig:setcompare}, only the Set 3 has a low rate of similarities between opaque or legit raw data content (3.5\%) and between $P^T$ and $P^F$ raw data (6\%).
This indicates that Set 3 is more suited for our raw data representation.
\begin{table}[h!]
	\centering
	\resizebox{\columnwidth}{!}{
		\begin{tabular}{|c|c|c|}
			\hline 
			\rowcolor{cyan!10}\textbf{Raw data} & \textbf{Detection similarities} & \textbf{Deobfuscation similarities} \\ 
			\hline
			\rowcolor{black!10} \textbf{Set 1} & 24.94\% & 31.92\%  \\
			
			\textbf{Set 2} & 17.38\% & 26.62\% \\
			
			\rowcolor{black!10} \textbf{Set 3} & \textbf{\color{red}{3.5\%}} & \textbf{\color{red}{6\%}} \\ 
			\hline 
		\end{tabular}
	}
	\caption{Percentage of our raw data content similarities for each sets.}
	\label{fig:setcompare}
\end{table}

\subsubsection{Decision tree based models.}
Decision trees~\cite{Rokach:2014:DMD:2755359} predict the output by learning simple decision rules deduced from the training dataset. 
The internal nodes of a decision tree contain binary conditions based on input features vectors, whereas the leaves are associated with the class labels we want to predict. 
Decision trees are built recursively. 
The root node contains all the training instances and each internal node splits its training instances into two subsets according to a condition based on the input. 
Leaf nodes however represent a classification or decision on these training instances.
Different approaches exist for the splitting conditions of internal nodes~\cite{DBLP:books/lib/HastieTF09}.
However, one downside of decision tree models is \textit{over-fitting}~\cite{DBLP:journals/csur/Dietterich95} which may cause the creation of over-complex trees that do not generalize the data well.
In our case, the decision tree model is capable of identifying and deobfuscating an opaque predicate $\mathcal{O}(\phi)$. 
We choose to create two distinct models: a first one that evaluates the stealth of an opaque predicate and a second one to evaluate its resiliency, as presented in the following paragraphs.

\paragraph{\textbf{Model for stealth (detection).}}
The construction of a classifier consists in the definition of a mapping function $\mathrm{C_{f}} : \mathcal{D} \rightarrow [0, 1]$ that, given a document $d$ (\textit{i.e.} an input),
returns a \textit{class label}, which is represented by a number (here 0 or 1) that defines the category of $d$. 
Applied to the evaluation of opaque predicates stealthiness, the function can be seen as $\mathrm{C_{f}} : \mathcal{D} \rightarrow [$NORMAL, OPAQUE$]$.
In other words, given the term-frequency vector of a symbolic execution state $D$, from a function entry point to a predicate, our model mapping 
function $C_{f}$ will return two values: {NORMAL} or {OPAQUE}.
If a model is capable of detecting a predicate as opaque, we can assume that the transformation is not stealthy.

\paragraph{\textbf{Model for resiliency (deobfuscation).}}
In order to evaluate the resiliency of an opaque predicate, we construct a model with a different function as presented for the evaluation of stealthiness.
Indeed, our goal is to predict if an opaque predicate is of type $P^T$ or $P^F$, thus, the function $\mathrm{C_{f}} : \mathcal{D} \rightarrow [0, 1]$ in that 
context can be expressed as $\mathrm{C_{f}} : \mathcal{D} \rightarrow [$TRUE, FALSE$]$.

~\\
 The choice of the best suited classification algorithm is often made on accuracy but in our work we choose our model based on its \textit{transparency} to easily interpret our results.
Since many learning algorithms exist, the next section will present our experiments to select the best classification model for both detection and deobfuscation of opaque predicates. 

\section{Experiments}
\label{sec:experiments}	
In this section we present our study of efficient and accurate creation of classification models.
We start by introducing the datasets variety used in our work.

\subsection{Datasets}
\label{subsec:datasets}
Our experiments are made on several C code samples. 
We use the scikit-learn API~\cite{scikit-learn} for the implementation of the models.
The datasets contain various types of code, each of them having different functionalities in order to have a model that does not fit to a specific type of program, as listed below:
~\\
\begin{itemize}
	\item GNU core utilities (\textit{i.e.} core-utils) binaries~\cite{Coreutils} for normal predicate samples;
	\item Cryptographic binaries for obfuscated and non-obfuscated predicates~\cite{Bcon};
	\item Samples from~\cite{DBLP:conf/acsac/BanescuCGNP16} containing basic algorithms (\textit{e.g.} factorial, sorting, etc.), non-cryptographic hash functions, small programs generated by Tigress;
	\item Samples involving the uses of structures and aliases~\cite{fragglet, thealgo}.
\end{itemize}
~\\
Our choice is motivated by their low ratio of dependencies and their straightforward compilation.
This makes their obfuscation possible using tools such as Tigress and OLLVM without errors during compilation.
\maria{A list of all different combinations of obfuscation transformations and options related to Tigress is given in Appendix \ref{appendix_tigress} and Listing \ref{lst:app_listing_tigress}.}

\paragraph{\textbf{Dataset size determination:}}
One important point is to determine the amount of samples required since this can significantly impact the cost of our studies and evaluations, as well as the reliability of our results.
If too much samples are collected, we face a longer evaluation time but if there are not enough samples is our dataset, our results may be irrelevant.
Several propositions based on statistical tests allow to determine the size of our datasets depending on the area of research~\cite{DBLP:journals/midm/FigueroaZKN12}.
Based on these works, we create our datasets with between 5000 and 15.000 samples in order to have a high probability of detection and of confidence level.
Each of our datasets are \emph{balanced}, \textit{i.e.} with an equal number of samples of each classes.
Next, we present our studies using these datasets.

\subsection{Preliminary studies}
The goal of our experiments is to investigate and answer the following questions:
~\\
\begin{itemize}
	\item \textbf{Study 1:} Which raw data language is the more efficient (in terms of time and space) and also the more accurate?
	\item \textbf{Study 2:} Which raw data content best expresses the normal and opaque predicates?
	\item \textbf{Study 3 and 4:} Which classification model is more accurate and which feature extraction algorithm is best suited?
\end{itemize}
~\\
The following paragraphs present our experiments for each question.
For this section and for our evaluations (see Section \ref{sec:evals}) we used a  laptop running Windows 7 with 16 GB of RAM and a Intel Core i7-6820HQ vPro processor.

\paragraph{\textbf{Study 1: Raw data language selection.}}
\label{subsub:s1}
Our goal is to select the most appropriate language for the symbolic execution engine.
We use Miasm-IR, which we compare with the translators it implements in SMT-LIBv2 language, C, and Python.
After normalizing these languages, as presented in Section \ref{subsec:raw_data}, we use our dataset of normal predicates from core-utils binaries along with structured-based opaque predicates from Tigress to study several points:
~\\
\begin{enumerate}
	\item Which set of samples is more efficient in terms of disk space?
	\item Which set of samples is more efficient in terms of computation time?
	\item Which language is more accurate for our models when representing our raw data?
\end{enumerate}
~\\
Table \ref{tab:comp_lang} illustrates our experiments using 20-fold cross-validation on decision-tree based models.
For each language, we used a dataset of 10000 balanced samples.

\begin{table}[h!]
	\centering
	\resizebox{\columnwidth}{!}{
		\begin{tabular}{|c|c|c|c|c|}
			\hline 
			\rowcolor{cyan!20}\textbf{Raw data language} & \textbf{Miasm2} & \textbf{SMT-LIBv2} & \textbf{C} & \textbf{Python} \\ 
			\hline 
			\rowcolor{black!10}Detection accuracy (\%) & \color{red}{94\%} & 90\% & 87\% & 87\% \\ 
			
			Deobfuscation accuracy (\%) & \color{red}{88\%} & 80\% & 78\% & 78\% \\ 
			
			\rowcolor{black!10}Execution time for detection (s) & \color{red}{15s} & 114s & 21s & 20s \\ 
			
			Execution time for deobfuscation (s) & \color{red}{12s} & 50s & 15s & 13s \\ 
			
			\rowcolor{black!10}Size of dataset (GB) & \color{red}{1.91GB} & 37.4GB & 2.11GB & 1.98GB \\ 
			\hline 
		\end{tabular}
	}
	\caption{Study of the raw data language accuracy and efficiency}
	\label{tab:comp_lang}
\end{table}

We observe that Miasm2 intermediate representation gives higher accuracy rates for both the detection and deobfuscation model. 
Moreover, it is more efficient in terms of disk space used (as opposed to the SMT-LIBv2 dataset), which leads to a faster time of execution. This is mainly due to the fact that Miasm2 intermediate language has a small set of terms expressing the semantics of the code as compared to other languages in our study.
According to these results, we choose Miasm2 for all of our raw data samples for the remaining of the paper.

\paragraph{\textbf{Study 2: Raw data content selection.}}
\label{subsub:s2}
It remains to single out the most suitable content that will express the construction of normal and invariant opaque predicates.
Table \ref{fig:setcompare} in Section \ref{subsec:raw_data} shows that the use of full symbolic state representation prevents having similarities between samples of different classes (\textit{i.e.} labels).
\begin{figure}[h!]
	\centering

	\begin{tikzpicture}
	\begin{axis}[
	ybar, axis on top,
	height=6cm, width=9cm,
	bar width=0.4cm,
	ymajorgrids, tick align=inside,
	major grid style={draw=black},
	enlarge y limits={value=.1,upper},
	ymin=0, ymax=100,
	axis x line*=bottom,
	axis y line*=left,
	y axis line style={opacity=100},
	tickwidth=0pt,
	enlarge x limits=true,
	legend style={
		at={(0.5,-0.2)},
		anchor=north,
		legend columns=-1
	},
	ylabel={Accuracy (\%)},
	symbolic x coords={
		Set-1, Set-2, Set-3},
	xtick=data,
	nodes near coords={
		\pgfmathprintnumber[precision=0]{\pgfplotspointmeta}
	}
	]
	\addplot [draw=none, fill=blue!50] coordinates {
		(Set-1, 77)
		(Set-2, 87)
		(Set-3, 94)
	};
	\addplot [draw=none,fill=red!50] coordinates {
		(Set-1, 57)
		(Set-2, 68)
		(Set-3, 88)
	};
	
	\legend{Detection, Deobfuscation}
	\end{axis}
	\end{tikzpicture}
	\caption{Predictions accuracy on the different raw data sets}
	\label{fig:s2}
\end{figure}
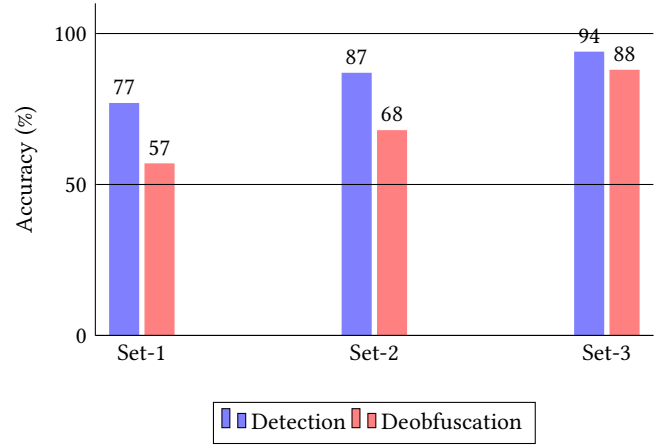
Thus, based on the same dataset of core-utils and structured-based opaque predicates generated with Tigress, we measure the average of our models accuracies for 
both detection and deobfuscation, evaluated with a 20-fold cross-validation.	
Figure \ref{fig:s2} confirms that the Set 3, \textit{i.e.} the full symbolic state, presents a better accuracy for both detection (at 94\%) and deobfuscation (at 88\%) when
using the decision tree algorithm on balanced datasets of 10000 samples.

\paragraph{\textbf{Study 3 and 4: Classification algorithm and feature extraction selection.}}
\label{subsub:s34}
In order to properly evaluate our methodology, we need to select the appropriate features extraction techniques combined with an accurate classification algorithm.

We have done experiments with the most common classifications models~\cite{DBLP:journals/informaticaSI/Kotsiantis07}, namely decision trees, k-nearest neighbors, support vector machines, neural network, naive Bayes, and random forest.
The use-case of our experiments is to evaluate the stealth of structured-based opaque predicates generated with Tigress on our datasets.
The features are expressed using \textit{term-frequency} (\textit{i.e.} tf) vectors as well as \textit{td-idf} vectors in order to compare both extraction techniques. Default parameters are applied for each classification algorithms used in our study.

\begin{table}[h!]
	\centering
	\resizebox{\columnwidth}{!}{
		\begin{tabular}{|c|c|c|}
			\hline 
			\rowcolor{cyan!10}\textbf{Classification algorithm} & \textbf{Term-frequency vectors} & \textbf{TD-IDF vectors} \\ 
			\hline 
			\rowcolor{black!10}\textbf{Decision-tree} & \textbf{\color{red}{94\%}} & \textbf{\color{red}{93\%}} \\
			\textbf{$k$-Nearest Neighbors} & 91\% & 92\% \\
			\rowcolor{black!10}\textbf{Support Vector Machine} & 87\% & 71\% \\
			\textbf{Linear Support Vector Machine} & 77\% & 83\% \\
			\rowcolor{black!10}\textbf{Multi-layer Perceptron} & 84\% & 92\% \\
			\textbf{Multinomial Naive-Bayes} & 58\% & 75\% \\
			\hline 
		\end{tabular}
	}
	\caption{Accuracies of different classification models using tf and td-idf vectors.}
	\label{tab:comparemodelsft}
\end{table}

Table \ref{tab:comparemodelsft} illustrate our results. 
We can observe that the decision tree model stands out from others when term-frequency vectors are used. It averages 94\% of detection accuracy whereas $k$-Nearest Neighbors averages 91\%.
As for the use of \textit{td-idf} vectors, the decision tree model has a better accuracy at 93\%.
\emph{According to this experiment, we choose the \textbf{Decision-tree} classification algorithm with \textbf{term-frequency} as features extraction technique in our methodology.}

\section{Evaluations}
\label{sec:evals}
Our goal in this section is to evaluate opaque predicates stealth and resiliency using a model based on decision trees.
We divide our evaluation into two parts:
~\\
\begin{enumerate}
	\item\emph{Stealth:} can the model differentiate an opaque predicate from a normal predicate, \textbf{\textit{i.e.} is the opaque predicate stealthy?}
	\item\emph{Resilience:} can the model differentiate a $P^T$ opaque predicate from a $P^F$ opaque predicate, \textbf{\textit{i.e.} is the opaque predicate resilient?}
\end{enumerate}

\subsection{Measuring stealth}
In this section we focus on the evaluation of stealthiness of opaque predicates. 
Namely, we want to see if our model is able to distinguish opaque predicates from normal predicates. 
Based on our datasets, our goal is to measure the efficiency of our model for the detection of opaque predicates based on different constructions.
\textit{Note that each datasets is balanced and contains 10000 samples.}

\paragraph{\textbf{Tigress:}}
The Tigress obfuscator can generate a variety of complex obfuscation transformations, \textit{e.g.} MBA-based, structured-based or environment-based.
To this end, we use several datasets of different opaque predicates constructions, balanced with normal predicates, to evaluate our model for detection. 
\textit{Dataset 1} contains arithmetic, MBA and environment-based opaque predicates whereas \textit{Dataset 2} contains 
structured-based (\textit{i.e.} alias-based) opaque predicates.
Moreover, we used a third dataset (\textit{Dataset 3}) that combines these opaque predicates with other obfuscation transformations such as arithmetic, 
literal, and data encodings (\textit{i.e.} \textit{EncA}, \textit{EncL}, and \textit{EncD}, respectively) joined with control-flow flattening (\textit{Flat})
and code virtualization (\textit{Virt}).
\begin{table}[ht!]
	\centering

	\resizebox{\columnwidth}{!}{
		\begin{tabular}{|c|c|c|c|c|c|}
			\hline 
			\rowcolor{cyan!10}\textbf{Datasets}& \textbf{Types of OP}&\textbf{Other transforms}&\textbf{Analysis time} &\textbf{Accuracy(\%)} &\textbf{F1 Score(\%)}\\ 
			\hline 		
			\textbf{Dataset 1} & \makecell{Arithmetic,\\ Environment-based} & None&1.13 s&93 \%&93 \%\\	
			\rowcolor{black!10}\textbf{Dataset 2} &{\colorbox{black!10}{\makecell{Arithmetic,\\ Structure-based}}} &None&2.14 s& 95 \%&95 \%\\
			\textbf{Dataset 3} &\makecell{Arithmetic, MBA,\\ Structure-based} & \makecell{EncA, EncL, EncD,\\ Flat, Virt}&1 s&99 \%&98 \%\\
			\hline 
	\end{tabular}}
	\caption{Evaluations of stealth (detection) using Tigress}
	\label{tab:tig_detect}
\end{table}
Our results are illustrated in Table \ref{tab:tig_detect}. 
Regardless of their types and of the implication of other obfuscation transformations, our detection model is able to efficiently predict if a predicate is \textit{opaque} or \textit{normal}. 
Indeed, the detection of arithmetic and environment-based opaque predicates scores an accuracy and F1 score of 93\%, whereas arithmetic and structured-based opaque predicates are less stealthy for our model with scores up to 95\%. 
However, as more obfuscation techniques are combined with opaque predicates, our predictions accuracy and F1 score rises to respectively 99\% and 98\%. 
This is due to the fact that opaque predicates patterns, once combined with other combination of transforms, become more specific thus lower their stealthiness. In our case however, code virtualization (\textit{i.e.} Virt) is applied before opaque predicates, as illustrated in Appendix \ref{appendix_tigress}. The opposite, namely applying code virtualization after other transformations, is a limitation to our methodology since the generated opaque predicates will be virtualized, thus transformed into byte-code.

\paragraph{\textbf{OLLVM:}}
In order to evaluate our model against opaque predicates generated by OLLVM, we split our evaluations in two sets. 
The first set uses samples obfuscated only with opaque predicates (\textit{i.e.} the bogus control-flow transformations \textit{bcf}).
The second set uses samples obfuscated with opaque predicates combined with control-flow flattening and instructions substitutions (\textit{i.e.} \textit{fla} and \textit{sub}, respectively) to see if we can evaluate opaque predicates stealthiness when they are combined with others transformations.
\begin{table}[ht!]
	\centering
	\resizebox{\columnwidth}{!}{
		\begin{tabular}{|c|c|c|c|c|c|}
			\hline 
			\rowcolor{cyan!10}\textbf{Datasets}& \textbf{Types of OP}&\textbf{Other transforms}&\textbf{Analysis time} &\textbf{Accuracy(\%)} &\textbf{F1 Score(\%)}\\ 
			\hline 

			\textbf{Dataset 1} & Arithmetic-based & None&2 s&89 \%&89 \%\\
			
			\rowcolor{black!10}\textbf{Dataset 2} & Arithmetic-based &fla, sub&1 s& 95 \%&94 \%\\
			
			\hline 
	\end{tabular}}
	\caption{Evaluations of stealth using OLLVM}
	\label{tab:ollvm_detect}
\end{table}
Table \ref{tab:ollvm_detect} illustrates our results. 
In the second dataset, our model is able to efficiently detect the labels of most predicates. 
However, when opaque predicates are not combined with other obfuscation transformation, we observe a loss of efficiency, from 95\% to 89\% accuracy.
This indicates that OLLVM opaque constructions are stealthier than other constructs, thus our model cannot properly distinguish them from normal predicates. 
At best, it will requires more training samples for our model in order to have a better accuracy.
One reason for their stealthiness in regard to our model is the fact that OLLVM arithmetic opaque predicates are bloc-centric, with basic encodings, which may have similar patterns to normal predicates from hash functions or cryptographic codes in our datasets. 
However, when they are combined to the other transformations, their patterns become more specific and our model has better prediction results.

\paragraph{\textbf{Bi-opaque:}}
Several constructions exist for bi-opaque predicates, among which float-based (\textit{i.e.} using floating instructions) or symbolic-memory based.
We use their obfuscator based on the OLLVM framework to evaluate our detection model.
\begin{table}[ht!]
	\centering
	\resizebox{\columnwidth}{!}{
		\begin{tabular}{|c|c|c|c|c|c|}
			\hline 
			\rowcolor{cyan!10}\textbf{Datasets}& \textbf{Types of OP}&\textbf{Other transforms}&\textbf{Analysis time} &\textbf{Accuracy(\%)} &\textbf{F1 Score(\%)}\\ 
			\hline 
			
			\textbf{Dataset 1} & Floats & None&0.6 s&93 \%&93 \%\\
			
			\rowcolor{black!10}\textbf{Dataset 2} & Symbolic-memory &fla, sub&0.9 s& 98 \%&98 \%\\
			
			\hline 
	\end{tabular}}
	\caption{Evaluations of stealth using Bi-opaque predicates from~\cite{DBLP:conf/dsn/XuZKTL18}}
	\label{tab:biop_detect}
\end{table}
As we can see in Table \ref{tab:biop_detect}, our model is efficient at detecting bi-opaque predicates with 93\% accuracy for float-based constructs. 
Bi-opaque predicates are constructed based on the same patterns as OLLVM opaque predicates but using floating-point instructions and registers instead.
However, symbolic-memory based constructs rely on more specific patterns, thus allowing a better detection rate at 98\% accuracy and F1 score.

\subsection{Measuring resiliency}
Once a predicate is detected as being \textit{opaque}, our goal is to measure its resiliency. 
In other words, we want to know if our model is able to deobfuscate, \textit{i.e.} predict the output of the opaque predicate. 
Our evaluations are based on invariant opaque predicates, $P^T$ and $P^F$, generated using different constructions.

\paragraph{\textbf{Tigress:}}
The patterns between $P^T$ and $P^F$ are more difficult to predict since both predicates are opaque and generated using the same construction.
However, the underlying invariant properties render our models efficient towards their deobfuscation. 
\begin{table}[ht!]
	\centering

	\resizebox{\columnwidth}{!}{
		\begin{tabular}{|c|c|c|c|c|c|}
			\hline 
			\rowcolor{cyan!10}\textbf{Datasets}& \textbf{Types of OP}&\textbf{Other transforms}&\textbf{Analysis time} &\textbf{Accuracy(\%)} &\textbf{F1 Score(\%)}\\ 
			\hline 		
			\textbf{Dataset 1} & \makecell{Arithmetic,\\ Environment} & None&0.3 s&90 \%&91 \%\\	
			\rowcolor{black!10}\textbf{Dataset 2} &{\colorbox{black!10}{\makecell{Arithmetic,\\ Structure}}} &None&1 s& 88 \%&87 \%\\
			\textbf{Dataset 3} &\makecell{Arithmetic, MBA,\\ Structure} & \makecell{EncA, EncL, EncD,\\ Flat, Virt}&3 s&92 \%&92 \%\\
			\hline 
	\end{tabular}}
	\caption{Evaluations of resiliency (deobfuscation) using Tigress}
	\label{tab:tig_deobf}
\end{table}
Table \ref{tab:tig_deobf} shows our results. 
We can observe that our model is able to detect environment-based invariants with scores of 90\% accuracy and 91\% of F1 score on balanced datasets of 5000 samples.
For structure-based invariants, we get slightly lower results, with 88\% and 87\% of accuracy and F1 score.
This is due to the fact that structured-based invariants use aliasing, producing patterns which are less dissimilar than for environment-based opaque predicates.
However, our model has a better accuracy and F1 score (92\% for both) when other transformations are used.
Thus, we are able to efficiently and accurately predict the invariant value of opaque predicates generated with Tigress, regardless of their constructions, 
and of the combination of obfuscation transformations used.

\paragraph{\textbf{Bi-opaque, OLLVM, and Tigress:}}
Since OLLVM only produces $P^T$ opaque predicates, we choose to combine all available samples generated from our three evaluated obfuscators.
A first dataset is used to evaluate our deobfuscation models against normal predicates and opaque predicates generated without any other transformations.
A second dataset is used to combined opaque predicates with others existing transformations from these obfuscators.
\begin{table}[ht!]
	\centering

	\resizebox{\columnwidth}{!}{
		\begin{tabular}{|c|c|c|c|c|c|}
			\hline 
			\rowcolor{cyan!10}\textbf{Datasets}& \textbf{Types of OP}&\textbf{Other transforms}&\textbf{Analysis time} &\textbf{Accuracy(\%)} &\textbf{F1 Score(\%)}\\ 
			\hline 		
			\textbf{Dataset 1} & \makecell{Arithmetic, MBA,\\ Environment, Structure,\\Symbolic-memory, Floats} & None&1 s&92 \%&91 \%\\	
			\rowcolor{black!10}\textbf{Dataset 2} &{\colorbox{black!10}{\makecell{Arithmetic, MBA,\\ Environment, Structure,\\Symbolic-memory, Floats}}} & \makecell{fla, bcf,\\EncA, EncL, EncD,\\ Flat, Virt}&0.5 s&95 \%&95 \%\\
			\hline 
	\end{tabular}}
	\caption{Evaluations of resiliency using Bi-opaque, OLLVM, and Tigress}
	\label{tab:deobf_all}
\end{table}
\textit{Note that all datasets are balanced and contain 15000 samples.}
Our results in Table \ref{tab:deobf_all} show that our methodology is efficient against all patterns of opaque predicates from available obfuscators.
Our model is able to detect the invariant patterns of all the opaque predicate constructs with 92\% accuracy and 91\% F1 score.
Moreover, when these opaque predicates are combined with other obfuscation transformations, the scores rise up to 95\%.

\begin{table*}[ht!]
	\centering
	\begin{tabular}{|c|c|c|c|c|c|}
		\hline 
		\rowcolor{cyan!10}\textbf{Tool} & \textbf{Obfuscator} & \textbf{Obfuscation} & \textbf{OP detection rate \%} & \textbf{\#FP, \#FN} &\textbf{Errors} \\ 
		\hline 
		& OLLVM & bcf & 100\% & 1,0 & 0\\
		& OLLVM & bcf, sub & 100\% & 0,0 & 1\\
		&Bi-opaque & float & 100\% & 4,0 & 0\\
		&Bi-opaque & symbolic-memory & 75\% & 1,5 & 2\\
		&Tigress & Environment-based & 60\% & 1,8 & 0\\
		&Tigress & Structure-based & 25\% & 2,12 & 1\\
		\multirow{-7}{*}{DROP}&Tigress & MBA, struct & 10\% & 0,10 & 8  \\
		\hline
		\rowcolor{black!10} & OLLVM & bcf & 100\% & 0,0 & 0\\
		\rowcolor{black!10}&OLLVM & bcf, sub & 100\% & 0,0 & 0\\
		\rowcolor{black!10}&Bi-opaque & float & 92\% & 0,0 & 0\\
		\rowcolor{black!10}	&Bi-opaque & symbolic-memory & 100\% & 1,0 & 0\\
		\rowcolor{black!10}&Tigress & Environment-based & 88\% & 2,3 & 0\\
		\rowcolor{black!10}	&Tigress & Structure-based & 82\% & 1,4 & 0\\
		\rowcolor{black!10}	\multirow{-7}{*}{Our methodology}&Tigress & MBA, struct & 85\% & 2,2 & 0\\
		\hline
		& OLLVM & bcf & 100\% & 0,0 & 0\\
		& OLLVM & bcf, sub & 100\% & 0,0 & 0\\
		&Bi-opaque & float & 100\% & 0,0 & 0\\
		& Bi-opaque & symbolic-memory & 85\% & 0,3 & 0\\
		&Tigress & Environment-based & 88\% & 1,2 & 0\\
		& Tigress & Structure-based & 65\% & 1,7 & 0\\
		\multirow{-7}{*}{Miasm DSE} &Tigress & MBA, struct & 52\% & 2,10 & 6 \\
		\hline
	\end{tabular} 
	
	\caption{Comparisons of opaque predicates deobfuscation using machine learning vs. SMT-solver based analyses.}
	\label{fig:comparedeobf}
\end{table*}

\subsection{Deobfuscation methodology}
\label{subsec:deobf}
Our methodology can be used as an efficient deobfuscation technique, if it is based on an adequate dataset of training samples.
We developed our methodology as an experimental IDA~\cite{IDAPro} plug-in that detects directly on the disassembled binary any opaque predicates and deobfuscates them, if needed.
We will compare our results with existing opaque predicates deobfuscation tools based on SMT solvers and symbolic execution, such as DROP~\cite{IDADROP}.
The latter is an IDA Pro plug-in based on Angr, which uses static symbolic execution for the removal of invariant and contextual opaque predicates.
Meanwhile, for the dynamic symbolic execution, we use Miasm2 dynamic symbolic execution engine.
We employ several datasets of opaque predicates obfuscated with various constructions and transformations. 
Moreover, we remove all samples used in our evaluations datasets from our learning samples used to built our model.

Our invariant opaque predicates are generated mainly  from~\cite{DBLP:conf/acsac/BanescuCGNP16} and Table \ref{fig:comparedeobf} shows the results. 
For each deobfuscation tool we use several samples obfuscated by different obfuscators (\textit{c.f.} column Obfuscator) and obfuscation transformations (\textit{c.f.} Obfuscation).
Column "OP detection rate" indicates the percentage of removed opaque predicates, whereas column "\#FP, \#FN" shows the number of false positive and false negative results respectively. 
Finally column "Errors" indicates if an error occurred during the analysis, \textit{e.g.} lack of memory or a timeout.

We observe that, for a static analysis, our experimental plug-in performs better at removing opaque predicates with complex constructs such as the one generated by Tigress,
or the bi-opaque constructs.
We obtain better results than the experimental plug-in DROP, as well as a better rate than DSE-based techniques for most constructions of opaque predicates.

\section{Limitations and perspectives}
\label{sec:limitations}
Our experiments and evaluations underline the efficiency of decision tree models to detect and deobfuscate opaque predicates. 
The most important achievement of our technique is that it allows a generalization to most invariant opaque predicates constructions.
Next we enumerate the limitations of our method.

The first limitation is due to decision tree models and the switch between obfuscators.
Namely, we can observe that a model that learns from samples generated using one obfuscator, cannot efficiently fit to transformations of another obfuscator if they use different kinds of constructions. 
This also hinders our ability to detect new constructions of opaque predicates.

A second limitation comes from the use of static symbolic execution to generate the symbolic state as a raw data. Such process is part of the deobfuscation 
application of our methodology, and, as any static analysis, may be time consuming.
This explains the use of our thresholded static symbolic execution in order to prevent as much as possible issues such as path explosion.

Our work proposes a new application of machine learning techniques for the purpose of evaluating obfuscation transformations, and also for removing them in a static automated manner.
Our experimentations and evaluations, indicate that our design can be extended to other complex constructions of opaque predicates such as thread-based and hash-based constructs. 
Future work includes also a more in-depth study of obfuscation transforms combinations and options as well as the generation of deobfuscated program to report any good or bad behaviors (\textit{e.g.} crashes).

\section{Related work}
\label{sec:relatedwork}
Many binary analysis techniques are often based on pattern matching for either detecting plagiarism, or malicious behaviors. 
Recent studies show the efficiency of machine learning and deep learning techniques for the detection and classification of malwares, \textit{e.g.}~\cite{DBLP:journals/jcs/RieckTWH11}, which also implicates the detection of similar codes within the malwares samples.
More closely related to the obfuscation area, the work in~\cite{DBLP:conf/acsac/SalemB16} aims at recovering meta-data information using machine learning techniques.
Their goal is to detect the obfuscation transformation used in several protected binaries generated by Tigress. 
Their evaluations show that naive Bayes and decision tree models can be efficient at detecting obfuscation transformations using filtered instruction traces. 
However, their work focuses on the recovery of informations about the obfuscation techniques used, but it does not aim at deobfuscating.

Another work, \cite{DBLP:conf/acsac/BanescuCGNP16}, aims at predicting the resiliency of obfuscated code against symbolic execution attacks. 
They use machine learning to measure the ability of several different symbolic execution engines to run against various layers and combinations of obfuscation techniques. 
Nevertheless, machine learning is not primarily used to remove any obfuscation transforms.

To summarize, existing work shows that machine learning techniques are pertinent \textit{w.r.t.} of the classification or the detection of features within binary samples. 
However, to the best of our knowledge, no deobfuscation study and methodology exists regarding these techniques. 
For this reason, in this paper, we proposed an efficient way to evaluate both the stealth and the resilience of opaque predicates through several studies and experiments combining binary analysis technique and machine learning.

\section{Conclusion}
\label{sec:conclusion}
In this paper we applied machine learning techniques to the evaluation of opaque predicates. 
By introducing the different constructions of opaque predicates and the limitations from dynamic symbolic execution techniques and SMT solvers, 
we underlined the importance of studying other alternatives for generic evaluations of these transformations.

We proposed a new approach that bridges a thresholded static symbolic execution with machine learning classification to evaluate both the stealth and resilience of invariant opaque predicates constructions.
The use of static symbolic execution allows us to have a better code coverage and scalability, which combined with a machine learning model, permits a generic approach by discarding the use of SMT solvers.
Our studies illustrate that our choices conduct towards the implementation of an efficient and accurate evaluation framework against state of the art obfuscators.
We created two models for the evaluation of stealth and resiliency of state-of-the-art opaque predicates constructions, with results up to 99\% for detection and 95\% for deobfuscation.
Moreover, we extended our work to a deobfuscation plug-in and compared our results to other tools, showing the efficiency of machine learning for the deobfuscation 
of most invariant opaque predicates constructions.
As future work, we propose to extend machine learning techniques to the evaluation of other obfuscation transformations as well as a more in-depth study of deep learning techniques, which we envision to render promising results.

We believe that our work provides a new framework to evaluate opaque predicates transformations, as well as a new alternative towards their static and automated deobfuscation.

\begin{acks}
This work is supported by the French National Research Agency in the framework of the Investissements d'Avenir program (ANR-15-IDEX-02).
\end{acks}

\appendix
\section{Tigress commands}
\label{appendix_tigress}
In the followings, we list the combinations of obfuscation transformations used for our datasets, in their application order. Note that the combinations listed in \textit{italic} are considered as clean samples since they do not generate opaque predicates. 
\begin{itemize}
	\item AddOpaque (16 or 32 times)
	\item AddOpaque, EncodeLiterals
	\item \textit{EncodeLiterals} 
	\item AddOpaque, EncodeArithmetics
	\item EncodeArithmetics, AddOpaque
	\item \textit{EncodeArithmetics}
	\item AddOpaque, EncodeData
	\item EncodeData, AddOpaque
	\item \textit{EncodeData}
	\item AddOpaque, EncodeArithmetics, EncodeLiterals, EncodeData
	\item EncodeData, EncodeArithemtics, EncodeLiterals, AddOpaque
	\item AddOpaque, Flatten 
	\item Flatten, AddOpaque
	\item \textit{Flatten}
	\item \textit{Flatten, EncodeData, EncodeArithemtics, EncodeLiterals}
	\item Virtualize, AddOpaque
	\item \textit{Virtualize}
	\item \textit{Virtualize, EncodeData, EncodeArithemtics, EncodeLiterals}
	\item \textit{Virtualize, Flatten}
	\item Flatten, AddOpaque, EncodeData, EncodeArithemtics, EncodeLiterals
	\item Virtualize, AddOpaque, EncodeData, EncodeArithemtics, EncodeLiterals
	\item Virtualize, Flatten, AddOpaque, EncodeData, EncodeArithemtics, EncodeLiterals
\end{itemize}

\subsection{Commands options}
\begin{lstlisting}[language=Python,caption={Tigress commands for sample generation}\label{lst:app_listing_tigress}]
# AddOpaque options
tigress --Transform=InitEntropy --Transform=InitOpaque --InitOpaqueStructs=list,array,env --Functions=main --Transform=AddOpaque --Functions=${3} --AddOpaqueCount=${NUM} --AddOpaqueKinds=call,fake,true

# Flatten
tigress --Transform=Flatten --FlattenDispatch=switch,goto --Functions=${3}

# Virtualize
tigress --Transform=Virtualize --VirtualizeDispatch=switch,direct,ifnest,linear --Functions=${3}

# EncodeLiterals
tigress --Transform=EncodeLiterals --Functions=${3} --EncodeLiteralsKinds=integer

# EncodeArithmetics
tigress --Transform=EncodeArithmetic --Functions=${3} --EncodeLiteralsKinds=integer

# EncodeData
tigress --Transform=EncodeData --LocalVariables=${4} --EncodeDataCodecs=poly,xor,add --Functions=${3}
\end{lstlisting}

	\bibliographystyle{ACM-Reference-Format}
	\bibliography{biblio_clones}


\begin{thebibliography}{45}


\ifx \showCODEN    \undefined \def \showCODEN     #1{\unskip}     \fi
\ifx \showDOI      \undefined \def \showDOI       #1{#1}\fi
\ifx \showISBNx    \undefined \def \showISBNx     #1{\unskip}     \fi
\ifx \showISBNxiii \undefined \def \showISBNxiii  #1{\unskip}     \fi
\ifx \showISSN     \undefined \def \showISSN      #1{\unskip}     \fi
\ifx \showLCCN     \undefined \def \showLCCN      #1{\unskip}     \fi
\ifx \shownote     \undefined \def \shownote      #1{#1}          \fi
\ifx \showarticletitle \undefined \def \showarticletitle #1{#1}   \fi
\ifx \showURL      \undefined \def \showURL       {\relax}        \fi
\providecommand\bibfield[2]{#2}
\providecommand\bibinfo[2]{#2}
\providecommand\natexlab[1]{#1}
\providecommand\showeprint[2][]{arXiv:#2}

\bibitem[\protect\citeauthoryear{Saudel and Salwan}{SST}{2015}]%
        {SSTIC2015-Saudel-Salwan}
 \bibinfo{year}{2015}\natexlab{}.
\newblock \bibinfo{booktitle}{\emph{Triton: A Dynamic Symbolic Execution
  Framework}}. \bibinfo{publisher}{SSTIC}.
\newblock


\bibitem[\protect\citeauthoryear{Algorithms}{Algorithms}{[n. d.]}]%
        {thealgo}
\bibfield{author}{\bibinfo{person}{The Algorithms}.} \bibinfo{year}{[n.
  d.]}\natexlab{}.
\newblock \bibinfo{title}{{C}}.
\newblock \bibinfo{howpublished}{\url{https://github.com/TheAlgorithms/C/}}.
\newblock
\newblock
\shownote{[Online; accessed 30-01-2019].}


\bibitem[\protect\citeauthoryear{Baldoni, Coppa, D'Elia, Demetrescu, and
  Finocchi}{Baldoni et~al\mbox{.}}{2018}]%
        {DBLP:journals/csur/BaldoniCDDF18}
\bibfield{author}{\bibinfo{person}{Roberto Baldoni}, \bibinfo{person}{Emilio
  Coppa}, \bibinfo{person}{Daniele~Cono D'Elia}, \bibinfo{person}{Camil
  Demetrescu}, {and} \bibinfo{person}{Irene Finocchi}.}
  \bibinfo{year}{2018}\natexlab{}.
\newblock \showarticletitle{A Survey of Symbolic Execution Techniques}.
\newblock \bibinfo{journal}{\emph{{ACM} Comput. Surv.}} \bibinfo{volume}{51},
  \bibinfo{number}{3} (\bibinfo{year}{2018}), \bibinfo{pages}{50:1--50:39}.
\newblock
\urldef\tempurl%
\url{https://doi.org/10.1145/3182657}
\showDOI{\tempurl}


\bibitem[\protect\citeauthoryear{Banescu, Collberg, Ganesh, Newsham, and
  Pretschner}{Banescu et~al\mbox{.}}{2016}]%
        {DBLP:conf/acsac/BanescuCGNP16}
\bibfield{author}{\bibinfo{person}{Sebastian Banescu},
  \bibinfo{person}{Christian~S. Collberg}, \bibinfo{person}{Vijay Ganesh},
  \bibinfo{person}{Zack Newsham}, {and} \bibinfo{person}{Alexander
  Pretschner}.} \bibinfo{year}{2016}\natexlab{}.
\newblock \showarticletitle{Code obfuscation against symbolic execution
  attacks}. In \bibinfo{booktitle}{\emph{Proceedings of the 32nd Annual
  Conference on Computer Security Applications, {ACSAC} 2016, USA}}.
  \bibinfo{pages}{189--200}.
\newblock
\urldef\tempurl%
\url{http://dl.acm.org/citation.cfm?id=2991114}
\showURL{%
\tempurl}


\bibitem[\protect\citeauthoryear{Bardin, David, and Marion}{Bardin
  et~al\mbox{.}}{2017}]%
        {DBLP:conf/sp/BardinDM17}
\bibfield{author}{\bibinfo{person}{S{\'{e}}bastien Bardin},
  \bibinfo{person}{Robin David}, {and} \bibinfo{person}{Jean{-}Yves Marion}.}
  \bibinfo{year}{2017}\natexlab{}.
\newblock \showarticletitle{Backward-Bounded {DSE:} Targeting Infeasibility
  Questions on Obfuscated Codes}. In \bibinfo{booktitle}{\emph{2017 {IEEE}
  Symposium on Security and Privacy, {SP} 2017, USA}}.
  \bibinfo{pages}{633--651}.
\newblock
\urldef\tempurl%
\url{https://doi.org/10.1109/SP.2017.36}
\showDOI{\tempurl}


\bibitem[\protect\citeauthoryear{Barrett, Fontaine, and Tinelli}{Barrett
  et~al\mbox{.}}{2017}]%
        {BarFT-RR-17}
\bibfield{author}{\bibinfo{person}{Clark Barrett}, \bibinfo{person}{Pascal
  Fontaine}, {and} \bibinfo{person}{Cesare Tinelli}.}
  \bibinfo{year}{2017}\natexlab{}.
\newblock \bibinfo{booktitle}{\emph{{The SMT-LIB Standard: Version 2.6}}}.
\newblock \bibinfo{type}{{T}echnical {R}eport}.
  \bibinfo{institution}{Department of Computer Science, The University of
  Iowa}.
\newblock
\newblock
\shownote{Available at {\tt www.SMT-LIB.org}.}


\bibitem[\protect\citeauthoryear{Biondi, Josse, Legay, and Sirvent}{Biondi
  et~al\mbox{.}}{2017}]%
        {DBLP:journals/compsec/BiondiJLS17}
\bibfield{author}{\bibinfo{person}{Fabrizio Biondi},
  \bibinfo{person}{S{\'{e}}bastien Josse}, \bibinfo{person}{Axel Legay}, {and}
  \bibinfo{person}{Thomas Sirvent}.} \bibinfo{year}{2017}\natexlab{}.
\newblock \showarticletitle{Effectiveness of synthesis in concolic
  deobfuscation}.
\newblock \bibinfo{journal}{\emph{Computers {\&} Security}}
  \bibinfo{volume}{70} (\bibinfo{year}{2017}), \bibinfo{pages}{500--515}.
\newblock
\urldef\tempurl%
\url{https://doi.org/10.1016/j.cose.2017.07.006}
\showDOI{\tempurl}


\bibitem[\protect\citeauthoryear{Blazytko, Contag, Aschermann, and
  Holz}{Blazytko et~al\mbox{.}}{2017}]%
        {DBLP:conf/uss/BlazytkoCAH17}
\bibfield{author}{\bibinfo{person}{Tim Blazytko}, \bibinfo{person}{Moritz
  Contag}, \bibinfo{person}{Cornelius Aschermann}, {and}
  \bibinfo{person}{Thorsten Holz}.} \bibinfo{year}{2017}\natexlab{}.
\newblock \showarticletitle{Syntia: Synthesizing the Semantics of Obfuscated
  Code}. In \bibinfo{booktitle}{\emph{26th {USENIX} Security Symposium,
  {USENIX} Security 2017, Canada}}. \bibinfo{pages}{643--659}.
\newblock
\urldef\tempurl%
\url{https://www.usenix.org/conference/usenixsecurity17/technical-sessions/presentation/blazytko}
\showURL{%
\tempurl}


\bibitem[\protect\citeauthoryear{Bruschi, Martignoni, and Monga}{Bruschi
  et~al\mbox{.}}{2006}]%
        {DBLP:conf/dimva/BruschiMM06}
\bibfield{author}{\bibinfo{person}{Danilo Bruschi}, \bibinfo{person}{Lorenzo
  Martignoni}, {and} \bibinfo{person}{Mattia Monga}.}
  \bibinfo{year}{2006}\natexlab{}.
\newblock \showarticletitle{Detecting Self-mutating Malware Using Control-Flow
  Graph Matching}. In \bibinfo{booktitle}{\emph{Detection of Intrusions and
  Malware {\&} Vulnerability Assessment, Third International Conference,
  {DIMVA} 2006 Proceedings, Germany}}. \bibinfo{pages}{129--143}.
\newblock
\urldef\tempurl%
\url{https://doi.org/10.1007/11790754\_8}
\showDOI{\tempurl}


\bibitem[\protect\citeauthoryear{Collberg, Martin, Myers, Zimmerman, Krajca,
  Kerneis, Debray, and Yadegari}{Collberg et~al\mbox{.}}{[n. d.]}]%
        {Tigress}
\bibfield{author}{\bibinfo{person}{Christian Collberg}, \bibinfo{person}{Sam
  Martin}, \bibinfo{person}{Jonathan Myers}, \bibinfo{person}{Bill Zimmerman},
  \bibinfo{person}{Petr Krajca}, \bibinfo{person}{Gabriel Kerneis},
  \bibinfo{person}{Saumya Debray}, {and} \bibinfo{person}{Babak Yadegari}.}
  \bibinfo{year}{[n. d.]}\natexlab{}.
\newblock \bibinfo{title}{{The Tigress C Diversifier/Obfuscator}}.
\newblock
  \bibinfo{howpublished}{\url{http://tigress.cs.arizona.edu/index.html}}.
\newblock
\newblock
\shownote{[Online; accessed 30-01-2019].}


\bibitem[\protect\citeauthoryear{Collberg, Thomborson, and Low}{Collberg
  et~al\mbox{.}}{1997}]%
        {Ctaxonomy}
\bibfield{author}{\bibinfo{person}{Christian Collberg}, \bibinfo{person}{Clark
  Thomborson}, {and} \bibinfo{person}{Douglas Low}.}
  \bibinfo{year}{1997}\natexlab{}.
\newblock \bibinfo{title}{A Taxonomy of Obfuscating Transformations}.
\newblock
\newblock


\bibitem[\protect\citeauthoryear{Collberg, Thomborson, and Low}{Collberg
  et~al\mbox{.}}{1998}]%
        {CollbergTL98}
\bibfield{author}{\bibinfo{person}{Christian~S. Collberg},
  \bibinfo{person}{Clark~D. Thomborson}, {and} \bibinfo{person}{Douglas Low}.}
  \bibinfo{year}{1998}\natexlab{}.
\newblock \showarticletitle{Manufacturing Cheap, Resilient, and Stealthy Opaque
  Constructs}. In \bibinfo{booktitle}{\emph{{POPL} '98, USA}}.
  \bibinfo{pages}{184--196}.
\newblock
\urldef\tempurl%
\url{https://doi.org/10.1145/268946.268962}
\showDOI{\tempurl}


\bibitem[\protect\citeauthoryear{Conte}{Conte}{[n. d.]}]%
        {Bcon}
\bibfield{author}{\bibinfo{person}{Brad Conte}.} \bibinfo{year}{[n.
  d.]}\natexlab{}.
\newblock \bibinfo{title}{{crypto-algorithms}}.
\newblock
  \bibinfo{howpublished}{\url{https://github.com/B-Con/crypto-algorithms}}.
\newblock
\newblock
\shownote{[Online; accessed 30-01-2019].}


\bibitem[\protect\citeauthoryear{Desclaux}{Desclaux}{2012}]%
        {Miasm}
\bibfield{author}{\bibinfo{person}{Fabrice Desclaux}.}
  \bibinfo{year}{2012}\natexlab{}.
\newblock \bibinfo{title}{Miasm : Framework de reverse engineering}.
\newblock \bibinfo{howpublished}{\url{https://github.com/cea-sec/miasm}}.
\newblock
\newblock
\shownote{[Online; accessed 30-01-2019].}


\bibitem[\protect\citeauthoryear{Dietterich}{Dietterich}{1995}]%
        {DBLP:journals/csur/Dietterich95}
\bibfield{author}{\bibinfo{person}{Thomas~G. Dietterich}.}
  \bibinfo{year}{1995}\natexlab{}.
\newblock \showarticletitle{Overfitting and Undercomputing in Machine
  Learning}.
\newblock \bibinfo{journal}{\emph{{ACM} Comput. Surv.}} \bibinfo{volume}{27},
  \bibinfo{number}{3} (\bibinfo{year}{1995}), \bibinfo{pages}{326--327}.
\newblock
\urldef\tempurl%
\url{https://doi.org/10.1145/212094.212114}
\showDOI{\tempurl}


\bibitem[\protect\citeauthoryear{Eyrolles, Goubin, and Videau}{Eyrolles
  et~al\mbox{.}}{2016}]%
        {DBLP:conf/ccs/EyrollesGV16}
\bibfield{author}{\bibinfo{person}{Ninon Eyrolles}, \bibinfo{person}{Louis
  Goubin}, {and} \bibinfo{person}{Marion Videau}.}
  \bibinfo{year}{2016}\natexlab{}.
\newblock \showarticletitle{Defeating MBA-based Obfuscation}. In
  \bibinfo{booktitle}{\emph{Proceedings of the 2016 {ACM} Workshop on Software
  PROtection, SPRO@CCS 2016, Austria}}. \bibinfo{pages}{27--38}.
\newblock
\urldef\tempurl%
\url{https://doi.org/10.1145/2995306.2995308}
\showDOI{\tempurl}


\bibitem[\protect\citeauthoryear{Figueroa, Zeng{-}Treitler, Kandula, and
  Ngo}{Figueroa et~al\mbox{.}}{2012}]%
        {DBLP:journals/midm/FigueroaZKN12}
\bibfield{author}{\bibinfo{person}{Rosa~L. Figueroa}, \bibinfo{person}{Qing
  Zeng{-}Treitler}, \bibinfo{person}{Sasikiran Kandula}, {and}
  \bibinfo{person}{Long~H. Ngo}.} \bibinfo{year}{2012}\natexlab{}.
\newblock \showarticletitle{Predicting sample size required for classification
  performance}.
\newblock \bibinfo{journal}{\emph{{BMC} Med. Inf. {\&} Decision Making}}
  \bibinfo{volume}{12} (\bibinfo{year}{2012}), \bibinfo{pages}{8}.
\newblock


\bibitem[\protect\citeauthoryear{Hastie, Tibshirani, and Friedman}{Hastie
  et~al\mbox{.}}{2009}]%
        {DBLP:books/lib/HastieTF09}
\bibfield{author}{\bibinfo{person}{Trevor Hastie}, \bibinfo{person}{Robert
  Tibshirani}, {and} \bibinfo{person}{Jerome~H. Friedman}.}
  \bibinfo{year}{2009}\natexlab{}.
\newblock \bibinfo{booktitle}{\emph{The elements of statistical learning: data
  mining, inference, and prediction, 2nd Edition}}.
\newblock \bibinfo{publisher}{Springer}.
\newblock
\showISBNx{9780387848570}
\urldef\tempurl%
\url{http://www.worldcat.org/oclc/300478243}
\showURL{%
\tempurl}


\bibitem[\protect\citeauthoryear{Hex-Rays}{Hex-Rays}{[n. d.]}]%
        {IDAPro}
\bibfield{author}{\bibinfo{person}{Hex-Rays}.} \bibinfo{year}{[n.
  d.]}\natexlab{}.
\newblock \bibinfo{title}{{IDA Pro : Interactive DisAssembler}}.
\newblock
  \bibinfo{howpublished}{\url{https://www.hex-rays.com/products/ida/index.shtml}}.
\newblock
\newblock
\shownote{[Online; accessed 30-01-2019].}


\bibitem[\protect\citeauthoryear{Howard}{Howard}{[n. d.]}]%
        {fragglet}
\bibfield{author}{\bibinfo{person}{Simon Howard}.} \bibinfo{year}{[n.
  d.]}\natexlab{}.
\newblock \bibinfo{title}{{c-algorithms}}.
\newblock
  \bibinfo{howpublished}{\url{https://github.com/fragglet/c-algorithms}}.
\newblock
\newblock
\shownote{[Online; accessed 30-01-2019].}


\bibitem[\protect\citeauthoryear{James}{James}{1985}]%
        {James:1985:CA:7557}
\bibfield{author}{\bibinfo{person}{Mike James}.}
  \bibinfo{year}{1985}\natexlab{}.
\newblock \bibinfo{booktitle}{\emph{Classification Algorithms}}.
\newblock \bibinfo{publisher}{Wiley-Interscience}, \bibinfo{address}{USA}.
\newblock
\showISBNx{0-471-84799-2}


\bibitem[\protect\citeauthoryear{Jones}{Jones}{2004}]%
        {DBLP:journals/jd/Jones04}
\bibfield{author}{\bibinfo{person}{Karen~Sp{\"{a}}rck Jones}.}
  \bibinfo{year}{2004}\natexlab{}.
\newblock \showarticletitle{A statistical interpretation of term specificity
  and its application in retrieval}.
\newblock \bibinfo{journal}{\emph{Journal of Documentation}}
  \bibinfo{volume}{60}, \bibinfo{number}{5} (\bibinfo{year}{2004}),
  \bibinfo{pages}{493--502}.
\newblock
\urldef\tempurl%
\url{https://doi.org/10.1108/00220410410560573}
\showDOI{\tempurl}


\bibitem[\protect\citeauthoryear{Junod, Rinaldini, Wehrli, and Michielin}{Junod
  et~al\mbox{.}}{2015}]%
        {ieeespro2015-JunodRWM}
\bibfield{author}{\bibinfo{person}{Pascal Junod}, \bibinfo{person}{Julien
  Rinaldini}, \bibinfo{person}{Johan Wehrli}, {and} \bibinfo{person}{Julie
  Michielin}.} \bibinfo{year}{2015}\natexlab{}.
\newblock \showarticletitle{Obfuscator-{LLVM} -- Software Protection for the
  Masses}. In \bibinfo{booktitle}{\emph{Proceedings of the {IEEE/ACM} 1st
  International Workshop on Software Protection, {SPRO'15}, Firenze, Italy, May
  19th, 2015}}, \bibfield{editor}{\bibinfo{person}{Brecht Wyseur}} (Ed.).
  \bibinfo{publisher}{IEEE}, \bibinfo{pages}{3--9}.
\newblock
\urldef\tempurl%
\url{https://doi.org/10.1109/SPRO.2015.10}
\showDOI{\tempurl}


\bibitem[\protect\citeauthoryear{Kernighan}{Kernighan}{1988}]%
        {Kernighan:1988:CPL:576122}
\bibfield{author}{\bibinfo{person}{Brian~W. Kernighan}.}
  \bibinfo{year}{1988}\natexlab{}.
\newblock \bibinfo{booktitle}{\emph{The C Programming Language}
  (\bibinfo{edition}{2nd} ed.)}.
\newblock \bibinfo{publisher}{Prentice Hall Professional Technical Reference}.
\newblock
\showISBNx{0131103709}


\bibitem[\protect\citeauthoryear{Kohavi}{Kohavi}{1995}]%
        {DBLP:conf/ijcai/Kohavi95}
\bibfield{author}{\bibinfo{person}{Ron Kohavi}.}
  \bibinfo{year}{1995}\natexlab{}.
\newblock \showarticletitle{A Study of Cross-Validation and Bootstrap for
  Accuracy Estimation and Model Selection}. In
  \bibinfo{booktitle}{\emph{Proceedings of the Fourteenth International Joint
  Conference on Artificial Intelligence, {IJCAI} 95, Canada}}.
  \bibinfo{pages}{1137--1145}.
\newblock
\urldef\tempurl%
\url{http://ijcai.org/Proceedings/95-2/Papers/016.pdf}
\showURL{%
\tempurl}


\bibitem[\protect\citeauthoryear{Kotsiantis}{Kotsiantis}{2007}]%
        {DBLP:journals/informaticaSI/Kotsiantis07}
\bibfield{author}{\bibinfo{person}{Sotiris~B. Kotsiantis}.}
  \bibinfo{year}{2007}\natexlab{}.
\newblock \showarticletitle{Supervised Machine Learning: {A} Review of
  Classification Techniques}.
\newblock \bibinfo{journal}{\emph{Informatica (Slovenia)}}
  \bibinfo{volume}{31}, \bibinfo{number}{3} (\bibinfo{year}{2007}),
  \bibinfo{pages}{249--268}.
\newblock
\urldef\tempurl%
\url{http://www.informatica.si/index.php/informatica/article/view/148}
\showURL{%
\tempurl}


\bibitem[\protect\citeauthoryear{Kovacheva}{Kovacheva}{2013}]%
        {DBLP:conf/iait/Kovacheva13}
\bibfield{author}{\bibinfo{person}{Aleksandrina Kovacheva}.}
  \bibinfo{year}{2013}\natexlab{}.
\newblock \showarticletitle{Efficient Code Obfuscation for Android}. In
  \bibinfo{booktitle}{\emph{Advances in Information Technology - 6th
  International Conference, {IAIT} 2013, Thailand.}} \bibinfo{pages}{104--119}.
\newblock
\urldef\tempurl%
\url{https://doi.org/10.1007/978-3-319-03783-7\_10}
\showDOI{\tempurl}


\bibitem[\protect\citeauthoryear{Lakhotia, Kumar, and Venable}{Lakhotia
  et~al\mbox{.}}{2005}]%
        {DBLP:journals/tse/LakhotiaKV05}
\bibfield{author}{\bibinfo{person}{Arun Lakhotia}, \bibinfo{person}{Eric~Uday
  Kumar}, {and} \bibinfo{person}{Michael Venable}.}
  \bibinfo{year}{2005}\natexlab{}.
\newblock \showarticletitle{A Method for Detecting Obfuscated Calls in
  Malicious Binaries}.
\newblock \bibinfo{journal}{\emph{{IEEE} Trans. Software Eng.}}
  \bibinfo{volume}{31}, \bibinfo{number}{11} (\bibinfo{year}{2005}),
  \bibinfo{pages}{955--968}.
\newblock
\urldef\tempurl%
\url{https://doi.org/10.1109/TSE.2005.120}
\showDOI{\tempurl}


\bibitem[\protect\citeauthoryear{Ming, Xu, Wang, and Wu}{Ming
  et~al\mbox{.}}{2015}]%
        {DBLP:conf/ccs/XWW15}
\bibfield{author}{\bibinfo{person}{Jiang Ming}, \bibinfo{person}{Dongpeng Xu},
  \bibinfo{person}{Li Wang}, {and} \bibinfo{person}{Dinghao Wu}.}
  \bibinfo{year}{2015}\natexlab{}.
\newblock \showarticletitle{{LOOP:} Logic-Oriented Opaque Predicate Detection
  in Obfuscated Binary Code}. In \bibinfo{booktitle}{\emph{Proceedings of the
  22nd {ACM} {SIGSAC} Conference on Computer and Communications Security, USA,
  October 12-6, 2015}}. \bibinfo{pages}{757--768}.
\newblock
\urldef\tempurl%
\url{https://doi.org/10.1145/2810103.2813617}
\showDOI{\tempurl}


\bibitem[\protect\citeauthoryear{Myles and Collberg}{Myles and
  Collberg}{2006}]%
        {DBLP:journals/ecr/MylesC06}
\bibfield{author}{\bibinfo{person}{Ginger Myles} {and}
  \bibinfo{person}{Christian~S. Collberg}.} \bibinfo{year}{2006}\natexlab{}.
\newblock \showarticletitle{Software watermarking via opaque predicates:
  Implementation, analysis, and attacks}.
\newblock \bibinfo{journal}{\emph{Electronic Commerce Research}}
  \bibinfo{volume}{6}, \bibinfo{number}{2} (\bibinfo{year}{2006}),
  \bibinfo{pages}{155--171}.
\newblock
\urldef\tempurl%
\url{https://doi.org/10.1007/s10660-006-6955-z}
\showDOI{\tempurl}


\bibitem[\protect\citeauthoryear{Pedregosa, Varoquaux, Gramfort, Michel,
  Thirion, Grisel, Blondel, Prettenhofer, Weiss, Dubourg, Vanderplas, Passos,
  Cournapeau, Brucher, Perrot, and Duchesnay}{Pedregosa et~al\mbox{.}}{2011}]%
        {scikit-learn}
\bibfield{author}{\bibinfo{person}{F. Pedregosa}, \bibinfo{person}{G.
  Varoquaux}, \bibinfo{person}{A. Gramfort}, \bibinfo{person}{V. Michel},
  \bibinfo{person}{B. Thirion}, \bibinfo{person}{O. Grisel},
  \bibinfo{person}{M. Blondel}, \bibinfo{person}{P. Prettenhofer},
  \bibinfo{person}{R. Weiss}, \bibinfo{person}{V. Dubourg}, \bibinfo{person}{J.
  Vanderplas}, \bibinfo{person}{A. Passos}, \bibinfo{person}{D. Cournapeau},
  \bibinfo{person}{M. Brucher}, \bibinfo{person}{M. Perrot}, {and}
  \bibinfo{person}{E. Duchesnay}.} \bibinfo{year}{2011}\natexlab{}.
\newblock \showarticletitle{Scikit-learn: Machine Learning in {P}ython}.
\newblock \bibinfo{journal}{\emph{Journal of Machine Learning Research}}
  \bibinfo{volume}{12} (\bibinfo{year}{2011}), \bibinfo{pages}{2825--2830}.
\newblock


\bibitem[\protect\citeauthoryear{Preda, Madou, Bosschere, and Giacobazzi}{Preda
  et~al\mbox{.}}{2006}]%
        {DBLP:conf/amast/PredaMBG06}
\bibfield{author}{\bibinfo{person}{Mila~Dalla Preda}, \bibinfo{person}{Matias
  Madou}, \bibinfo{person}{Koen~De Bosschere}, {and} \bibinfo{person}{Roberto
  Giacobazzi}.} \bibinfo{year}{2006}\natexlab{}.
\newblock \showarticletitle{Opaque Predicates Detection by Abstract
  Interpretation}. In \bibinfo{booktitle}{\emph{Algebraic Methodology and
  Software Technology, 11th International Conference, {AMAST} 2006, Estonia}}.
  \bibinfo{pages}{81--95}.
\newblock
\urldef\tempurl%
\url{https://doi.org/10.1007/11784180\_9}
\showDOI{\tempurl}


\bibitem[\protect\citeauthoryear{Project}{Project}{2002}]%
        {Coreutils}
\bibfield{author}{\bibinfo{person}{GNU Project}.}
  \bibinfo{year}{2002}\natexlab{}.
\newblock \bibinfo{title}{{GNU Core Utilities}}.
\newblock
  \bibinfo{howpublished}{\url{https://www.gnu.org/software/coreutils/}}.
\newblock
\newblock
\shownote{[Online; accessed 30-01-2019].}


\bibitem[\protect\citeauthoryear{Rieck, Trinius, Willems, and Holz}{Rieck
  et~al\mbox{.}}{2011}]%
        {DBLP:journals/jcs/RieckTWH11}
\bibfield{author}{\bibinfo{person}{Konrad Rieck}, \bibinfo{person}{Philipp
  Trinius}, \bibinfo{person}{Carsten Willems}, {and} \bibinfo{person}{Thorsten
  Holz}.} \bibinfo{year}{2011}\natexlab{}.
\newblock \showarticletitle{Automatic analysis of malware behavior using
  machine learning}.
\newblock \bibinfo{journal}{\emph{Journal of Computer Security}}
  \bibinfo{volume}{19}, \bibinfo{number}{4} (\bibinfo{year}{2011}),
  \bibinfo{pages}{639--668}.
\newblock


\bibitem[\protect\citeauthoryear{Rinsma}{Rinsma}{2017}]%
        {IDADROP}
\bibfield{author}{\bibinfo{person}{Thomas Rinsma}.}
  \bibinfo{year}{2017}\natexlab{}.
\newblock \bibinfo{title}{{Seeing through obfuscation: interactive detection
  and removal of opaque predicates}}.
\newblock
  \bibinfo{howpublished}{\url{https://github.com/Riscure/DROP-IDA-plugin}}.
\newblock
\newblock
\shownote{[Online; accessed 30-01-2019].}


\bibitem[\protect\citeauthoryear{Rokach and Maimon}{Rokach and Maimon}{2014}]%
        {Rokach:2014:DMD:2755359}
\bibfield{author}{\bibinfo{person}{Lior Rokach} {and} \bibinfo{person}{Oded
  Maimon}.} \bibinfo{year}{2014}\natexlab{}.
\newblock \bibinfo{booktitle}{\emph{Data Mining With Decision Trees: Theory and
  Applications} (\bibinfo{edition}{2nd} ed.)}.
\newblock \bibinfo{publisher}{World Scientific Publishing Co., Inc.},
  \bibinfo{address}{USA}.
\newblock
\showISBNx{9789814590075, 981459007X}


\bibitem[\protect\citeauthoryear{Rossum}{Rossum}{1995}]%
        {Rossum:1995:PRM:869369}
\bibfield{author}{\bibinfo{person}{Guido Rossum}.}
  \bibinfo{year}{1995}\natexlab{}.
\newblock \bibinfo{booktitle}{\emph{Python Reference Manual}}.
\newblock \bibinfo{type}{{T}echnical {R}eport}. \bibinfo{address}{Amsterdam,
  The Netherlands, The Netherlands}.
\newblock


\bibitem[\protect\citeauthoryear{Salem and Banescu}{Salem and Banescu}{2016}]%
        {DBLP:conf/acsac/SalemB16}
\bibfield{author}{\bibinfo{person}{Aleieldin Salem} {and}
  \bibinfo{person}{Sebastian Banescu}.} \bibinfo{year}{2016}\natexlab{}.
\newblock \showarticletitle{Metadata recovery from obfuscated programs using
  machine learning}. In \bibinfo{booktitle}{\emph{Proceedings of the 6th
  Workshop on Software Security, Protection, and Reverse Engineering, SSPREW
  2016, USA, 2016}}. \bibinfo{pages}{1:1--1:11}.
\newblock
\urldef\tempurl%
\url{https://doi.org/10.1145/3015135.3015136}
\showDOI{\tempurl}


\bibitem[\protect\citeauthoryear{Schrittwieser, Katzenbeisser, Kinder,
  Merzdovnik, and Weippl}{Schrittwieser et~al\mbox{.}}{2016}]%
        {DBLP:journals/csur/SchrittwieserKK16}
\bibfield{author}{\bibinfo{person}{Sebastian Schrittwieser},
  \bibinfo{person}{Stefan Katzenbeisser}, \bibinfo{person}{Johannes Kinder},
  \bibinfo{person}{Georg Merzdovnik}, {and} \bibinfo{person}{Edgar~R. Weippl}.}
  \bibinfo{year}{2016}\natexlab{}.
\newblock \showarticletitle{Protecting Software through Obfuscation: Can It
  Keep Pace with Progress in Code Analysis?}
\newblock \bibinfo{journal}{\emph{{ACM} Comput. Surv.}} \bibinfo{volume}{49},
  \bibinfo{number}{1} (\bibinfo{year}{2016}), \bibinfo{pages}{4:1--4:37}.
\newblock
\urldef\tempurl%
\url{https://doi.org/10.1145/2886012}
\showDOI{\tempurl}


\bibitem[\protect\citeauthoryear{Shoshitaishvili, Wang, Salls, Stephens,
  Polino, Dutcher, Grosen, Feng, Hauser, Kruegel, and Vigna}{Shoshitaishvili
  et~al\mbox{.}}{2016}]%
        {shoshitaishvili2016state}
\bibfield{author}{\bibinfo{person}{Yan Shoshitaishvili}, \bibinfo{person}{Ruoyu
  Wang}, \bibinfo{person}{Christopher Salls}, \bibinfo{person}{Nick Stephens},
  \bibinfo{person}{Mario Polino}, \bibinfo{person}{Audrey Dutcher},
  \bibinfo{person}{John Grosen}, \bibinfo{person}{Siji Feng},
  \bibinfo{person}{Christophe Hauser}, \bibinfo{person}{Christopher Kruegel},
  {and} \bibinfo{person}{Giovanni Vigna}.} \bibinfo{year}{2016}\natexlab{}.
\newblock \showarticletitle{{SoK: (State of) The Art of War: Offensive
  Techniques in Binary Analysis}}. In \bibinfo{booktitle}{\emph{IEEE Symposium
  on Security and Privacy}}.
\newblock


\bibitem[\protect\citeauthoryear{Sutter, Basile, Ceccato, Falcarin, Zunke,
  Wyseur, and d'Annoville}{Sutter et~al\mbox{.}}{2016}]%
        {DBLP:conf/ccs/SutterBCFZWD16}
\bibfield{author}{\bibinfo{person}{Bjorn~De Sutter}, \bibinfo{person}{Cataldo
  Basile}, \bibinfo{person}{Mariano Ceccato}, \bibinfo{person}{Paolo Falcarin},
  \bibinfo{person}{Michael Zunke}, \bibinfo{person}{Brecht Wyseur}, {and}
  \bibinfo{person}{J{\'{e}}r{\^{o}}me d'Annoville}.}
  \bibinfo{year}{2016}\natexlab{}.
\newblock \showarticletitle{The {ASPIRE} Framework for Software Protection}. In
  \bibinfo{booktitle}{\emph{Proceedings of the 2016 {ACM} Workshop on Software
  PROtection, SPRO@CCS 2016, Vienna, Austria, October 24-28, 2016}},
  \bibfield{editor}{\bibinfo{person}{Brecht Wyseur} {and}
  \bibinfo{person}{Bjorn~De Sutter}} (Eds.). \bibinfo{publisher}{{ACM}},
  \bibinfo{pages}{91--92}.
\newblock
\showISBNx{978-1-4503-4576-7}
\urldef\tempurl%
\url{https://doi.org/10.1145/2995306.2995316}
\showDOI{\tempurl}


\bibitem[\protect\citeauthoryear{Tofighi{-}Shirazi, Christofi, Elbaz{-}Vincent,
  and Le}{Tofighi{-}Shirazi et~al\mbox{.}}{2018}]%
        {DBLP:conf/acsac/Tofighi-Shirazi18}
\bibfield{author}{\bibinfo{person}{Ramtine Tofighi{-}Shirazi},
  \bibinfo{person}{Maria Christofi}, \bibinfo{person}{Philippe
  Elbaz{-}Vincent}, {and} \bibinfo{person}{Thanh~Ha Le}.}
  \bibinfo{year}{2018}\natexlab{}.
\newblock \showarticletitle{{DoSE: Deobfuscation based on Semantic
  Equivalence}}. In \bibinfo{booktitle}{\emph{Proceedings of the 8th Software
  Security, Protection, and Reverse Engineering Workshop, USA}}.
  \bibinfo{pages}{1:1--1:12}.
\newblock
\urldef\tempurl%
\url{https://doi.org/10.1145/3289239.3289243}
\showDOI{\tempurl}


\bibitem[\protect\citeauthoryear{Udupa, Debray, and Madou}{Udupa
  et~al\mbox{.}}{2005}]%
        {DBLP:conf/wcre/UdupaDM05}
\bibfield{author}{\bibinfo{person}{Sharath~K. Udupa},
  \bibinfo{person}{Saumya~K. Debray}, {and} \bibinfo{person}{Matias Madou}.}
  \bibinfo{year}{2005}\natexlab{}.
\newblock \showarticletitle{Deobfuscation: Reverse Engineering Obfuscated
  Code}. In \bibinfo{booktitle}{\emph{12th Working Conference on Reverse
  Engineering, {WCRE} 2005, USA}}. \bibinfo{pages}{45--54}.
\newblock
\urldef\tempurl%
\url{https://doi.org/10.1109/WCRE.2005.13}
\showDOI{\tempurl}


\bibitem[\protect\citeauthoryear{Xu, Zhou, Kang, Tu, and Lyu}{Xu
  et~al\mbox{.}}{2018}]%
        {DBLP:conf/dsn/XuZKTL18}
\bibfield{author}{\bibinfo{person}{Hui Xu}, \bibinfo{person}{Yangfan Zhou},
  \bibinfo{person}{Yu Kang}, \bibinfo{person}{Fengzhi Tu}, {and}
  \bibinfo{person}{Michael~R. Lyu}.} \bibinfo{year}{2018}\natexlab{}.
\newblock \showarticletitle{Manufacturing Resilient Bi-Opaque Predicates
  Against Symbolic Execution}. In \bibinfo{booktitle}{\emph{48th Annual
  {IEEE/IFIP} International Conference on Dependable Systems and Networks,
  {DSN} 2018, Luxembourg}}. \bibinfo{pages}{666--677}.
\newblock
\urldef\tempurl%
\url{https://doi.org/10.1109/DSN.2018.00073}
\showDOI{\tempurl}


\bibitem[\protect\citeauthoryear{Zhou, Main, Gu, and Johnson}{Zhou
  et~al\mbox{.}}{2007}]%
        {DBLP:conf/wisa/ZhouMGJ07}
\bibfield{author}{\bibinfo{person}{Yongxin Zhou}, \bibinfo{person}{Alec Main},
  \bibinfo{person}{Yuan~Xiang Gu}, {and} \bibinfo{person}{Harold Johnson}.}
  \bibinfo{year}{2007}\natexlab{}.
\newblock \showarticletitle{Information Hiding in Software with Mixed
  Boolean-Arithmetic Transforms}. In \bibinfo{booktitle}{\emph{Information
  Security Applications, 8th International Workshop, {WISA} 2007, Korea}}.
  \bibinfo{pages}{61--75}.
\newblock
\urldef\tempurl%
\url{https://doi.org/10.1007/978-3-540-77535-5\_5}
\showDOI{\tempurl}


\end{thebibliography}
	
\end{document}